\documentclass[apj]{emulateapj}

\def \nustar {\emph{NuSTAR}}

\begin{document}

\submitted{Accepted to the Astropysical Journal on December 16, 2014}

\shorttitle{A Multi-wavelength Study of PSR\,J1023+0038}
\shortauthors{Bogdanov et al.}

\title{Coordinated X-ray, Ultraviolet, Optical, and Radio Observations of the PSR J1023+0038 System in a Low-Mass X-ray Binary State}

\author{Slavko Bogdanov\altaffilmark{1}, Anne
  M.~Archibald\altaffilmark{2}, Cees Bassa\altaffilmark{2}, Adam T.~Deller\altaffilmark{2},  Jules
  P.~Halpern\altaffilmark{1}, George Heald\altaffilmark{2,3}, \\ Jason
  W.~T.~Hessels\altaffilmark{2,4}, Gemma H.~Janssen\altaffilmark{2}, Andrew G.~Lyne\altaffilmark{5}, Javier Mold\'on\altaffilmark{2},
  Zsolt Paragi\altaffilmark{6}, \\ Alessandro Patruno\altaffilmark{7,2},
  Benetge B.~P.~Perera\altaffilmark{5}, Ben W.~Stappers\altaffilmark{5},
  Shriharsh P.~Tendulkar\altaffilmark{8}, \\ Caroline R.~D'Angelo\altaffilmark{7}, Rudy Wijnands\altaffilmark{4}}

\altaffiltext{1}{Columbia Astrophysics Laboratory, Columbia University, 550 West 120th Street, New York, NY 10027, USA; slavko@astro.columbia.edu}

\altaffiltext{2}{ASTRON, the Netherlands Institute for Radio
  Astronomy, Postbus 2, 7990 AA, Dwingeloo, The Netherlands}

\altaffiltext{3}{Kapteyn Astronomical Institute, University of
  Groningen, PO Box 800, 9700 AV, Groningen, The Netherlands}

\altaffiltext{4}{Anton Pannekoek
Institute for Astronomy, University of Amsterdam, Science Park 904,
1098 XH Amsterdam, The Netherlands}

\altaffiltext{5}{Jodrell Bank Centre for Astrophysics, School of Physics and Astronomy, The University of Manchester, Manchester M13 9PL, UK}

\altaffiltext{6}{JIVE, Joint Institute for VLBI in Europe, Postbus 2, 7990 AA Dwingeloo, The Netherlands}

\altaffiltext{7}{Leiden Observatory, Leiden University, PO Box 9513, 2300 RA, Leiden, The Netherlands}

\altaffiltext{8}{California Institute of Technology, 1200 East
  California Boulevard, Pasadena, CA 91125, USA}

\begin{abstract}  
The PSR J1023+0038 binary system hosts a neutron star and a low-mass,
main-sequence-like star.  It switches on year timescales between
states as an eclipsing radio millisecond pulsar and a low-mass X-ray
binary. We present a multi-wavelength observational campaign of
PSR\,J1023+0038 in its most recent low-mass X-ray binary state.  Two
long \textit{XMM-Newton} observations reveal that the system spends
$\sim$70\% of the time in a $\approx$$3\times10^{33}$\,erg s$^{-1}$
X-ray luminosity mode, which, as shown in \citet{Arch14}, exhibits
coherent X-ray pulsations. This emission is interspersed with frequent
lower flux mode intervals with $\approx$$5\times 10^{32}$\,erg
s$^{-1}$ and sporadic flares reaching up to $\approx$$10^{34}$\,erg
s$^{-1}$, with neither mode showing significant X-ray pulsations. The
switches between the three flux modes occur on timescales of order 10
s.  In the UV and optical, we observe occasional intense flares
coincident with those observed in X-rays.  Our radio timing
observations reveal no pulsations at the pulsar period during any of
the three X-ray modes, presumably due to complete quenching of the
radio emission mechanism by the accretion flow.  Radio imaging detects
highly variable, flat-spectrum continuum radiation from PSR
J1023+0038, consistent with an origin in a weak jet-like
  outflow. Our concurrent X-ray and radio continuum data sets do not
exhibit any correlated behavior. The observational evidence we present
bears qualitative resemblance to the behavior predicted by some
existing ``propeller'' and ``trapped'' disk accretion models although
none can account for key aspects of the rich phenomenology of this
system.
\end{abstract}

\keywords{pulsars: general --- pulsars: individual (PSR J1023+0038) --- stars: neutron --- X-rays: binaries}

\section{INTRODUCTION}
PSR\,J1023+0038 (also known as AY Sextantis or FIRST J102347.6+003841)
is a $1.7$\,ms pulsar in a $4.75$\,hour binary orbit around a bloated
$\sim$0.2\,M$_{\odot}$ main-sequence-like companion star.  This system
is notable in that it was the first to exhibit compelling evidence for
the transition process between an accretion disk-dominated low-mass
X-ray binary-like state and a disk-free radio pulsar state. Optical
observations revealed an accretion disk in the system in 2001
\citep{Bond02,Szkody03,Wang09}, which seemed to be absent in early
2003 \citep{Thor05} and at the time of the radio pulsar discovery
\citep{Arch09}.  During its disk episode in 2001, the binary
appeared optically blue and bright in addition to showing strong,
double-peaked H and He emission lines, which are commonly seen in
low-mass X-ray binaries (LMXBs), and are a typical feature of
accretion disks. This suggests that the companion star overflowed its
Roche lobe, forming a disk surrounding the millisecond pulsar (MSP).
In contrast, spectra obtained starting in early 2003 showed a
substantially lower optical flux and a typical G-type spectrum,
implying the absence of a substantial accretion disk.

A recent torrent of developments have revealed two close analogs to
PSR\,J1023+0038. In 2013 March, PSR\,J1824--2452I in the globular
cluster M28 was seen to switch between rotation-powered (radio) and
high-luminosity accretion-powered (X-ray) pulsations \citep{Pap13},
thereby strenghtening the long-suspected evolutionary link between
LMXBs and ``recycled'' pulsars \citep{Alp82}.  In addition,
re-examination of archival optical and X-ray data of the X-ray binary
XSS\,J12270--4859 revealed that only a few months prior (in 2012
November/December) its accretion disk had disappeared and the 1.7-ms
radio pulsar became active \citep{Bassa14,Bog14b,Roy14}. See
\citet{Lin14b} for an overview of the X-ray states of these and
analogous objects.

%
%
\begin{figure*}[!t]
\begin{center}
\includegraphics[width=0.55\textwidth]{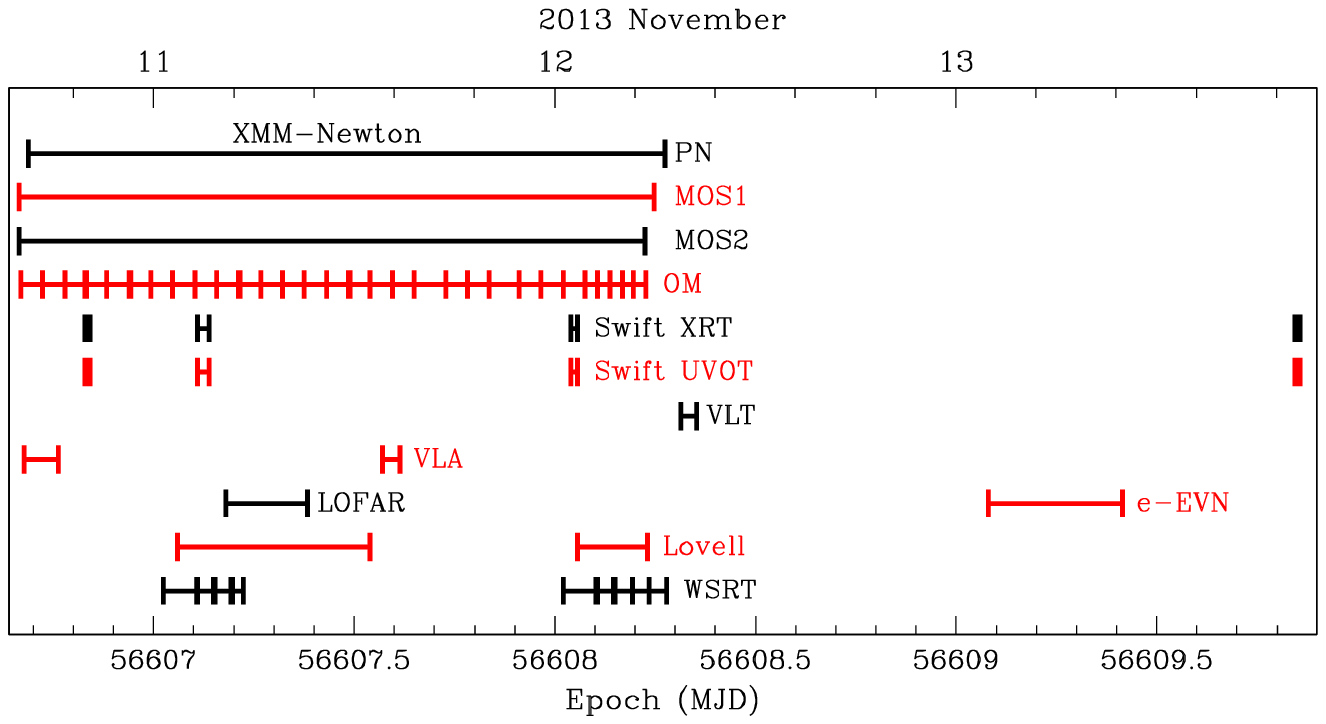}
\includegraphics[width=0.329\textwidth]{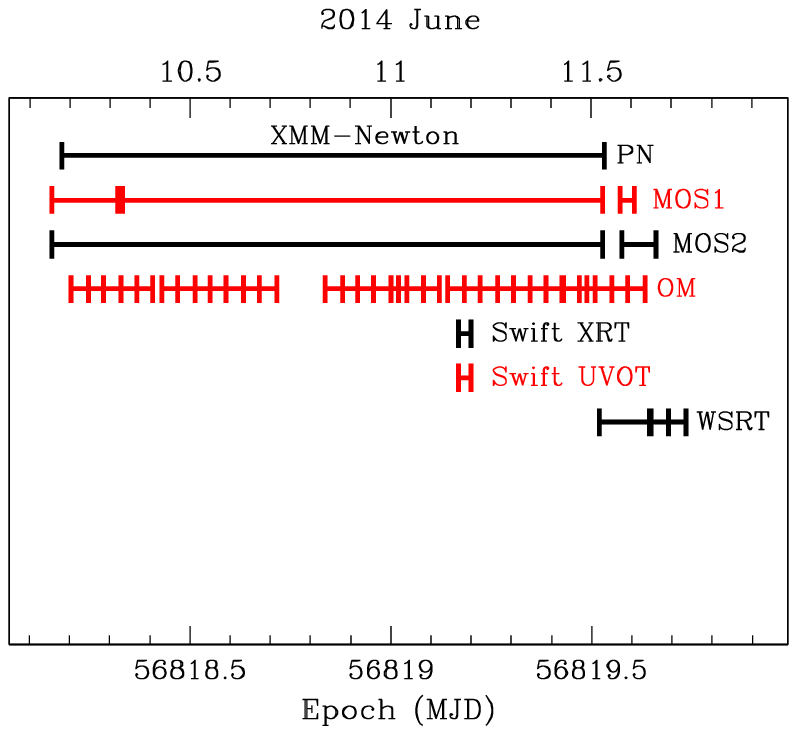}
\end{center}
\caption{Timeline of the contemporaneous or near-contemporaneous
  multi-wavelength observations presented in this paper. The color
  variation is provided for ease of interpretation.}
\end{figure*} 

Unexpectedly, on 2013 June 23, no radio pulsations were detected from
PSR\,J1023+0038, and none have been detected up to the
  publication of this article despite an intensified monitoring
campaign and higher-radio-frequency observations with greater
sensitivity. Subsequent X-ray, optical, and $\gamma$-ray investigation
revealed that PSR J1023+0038 has undergone another transformation to
an accretion disk-dominated state
\citep{Pat14,Stap14,Halpern13,Ten14}. This transition was accompanied
by an extraordinary five-fold increase in high-energy $\gamma$-ray luminosity
as seen by the \textit{Fermi} Large Area Telescope \citep{Stap14}.  The X-ray emission observed by \textit{Swift} XRT
revealed a significant change in behavior compared to the disk-free
state as well: in addition to an increase in flux by over an order of
magnitude to $\sim$$10^{33-34}$\,erg s$^{-1}$, the
orbital-phase-dependent modulations had disappeared, replaced by
aperiodic variability with rapid drops to a low flux level
\citep{Pat14}. Follow-up hard X-ray observations with \textit{NuSTAR}
revealed strong flares reaching up to $\approx$$1.2\times
10^{34}$\,erg s$^{-1}$ (3--79 keV), as well as the same peculiar drops
in flux \citep{Ten14}.  This behavior bears close resemblance to that
observed in PSR\,J1824--2452I \citep{Lin14a} and XSS\,J12270--4859
\citep{deM10,deM13} in their low-luminosity LMXB states.

In \citet{Arch14}, we established that despite PSR J1023+0038 existing
at a luminosity level typical of quiescent LMXBs,
channeled accretion onto the neutron star surface appears to
be occuring on a regular basis as demonstrated by the detection of
coherent X-ray pulsations. Herein, we build upon this finding with a
more in-depth analysis of the same \textit{XMM-Newton} data set, which
is augmented by an array of coordinated multi-wavelength observations
of PSR J1023+0038. This study provides the most complete observational
picture to date of the LMXB state of the PSR J1023+0038 system and,
arguably, transition millisecond pulsar systems in general.  The work
is organized as follows. In \S2, we summarize the observations and
data analysis. In \S3, we focus on the X-ray variability, while in \S4
we focus on its statistical properties. We discuss the X-ray
pulsations in \S5 and summarize the spectroscopic analysis of the
X-ray emission in \S6.  We describe the results of the optical
photomety and spectroscopy in \S7 and \S8. We present the results of
our radio imaging and timing observations in \S9 and \S10. Finally, in
\S11 we provide a discussion and offer conclusions in \S12.

\begin{deluxetable*}{lcrcc}
\tabletypesize{\footnotesize}
\tablewidth{0pt}
\tablecaption{Log of Observations of PSR J1023+0038 during 2013 November 10--12.}
\tablehead{
\colhead{Telescope/} &
\colhead{Start time} & \colhead{Duration} & \colhead{Band/} & \colhead{Mode}\\
\colhead{Instrument} & \colhead{(MJD)} & \colhead{(s)} & \colhead{Filter} & \colhead{}}
\startdata
\textit{XMM-Newton}/EPIC pn	&	56606.69084	&	136981	&	0.3--10 keV	&	Fast timing	\\
\textit{XMM-Newton}/EPIC MOS1	&	56606.66645	&	136942	&	0.3--10	keV &	Small Window	\\
\textit{XMM-Newton}/EPIC MOS2	&	56606.66696	&	134893	&	0.3--10	keV &	Small Window	\\
\hline
\textit{Swift}/XRT	&	56606.83056	&	1161	&	0.3--10\,keV &	Fast	\\
\nodata	&	56607.11389	&	2312	&	0.3--10\,keV &	Fast	\\
\nodata	&	56608.04375	&	1138	&	0.3--10\,keV &	Fast	\\
\nodata	&	56609.84792	&	1098	&	0.3--10\,keV &	Fast \\
\hline
\textit{XMM-Newton}/OM	&	56606.67079	&	4656	&	$B$	&	Image Fast	\\
\nodata	&	56606.72470	&	4720	&	$B$	&	Image Fast	\\
\nodata	&	56606.77935	&	4720	&	$B$	&	Image Fast	\\
\nodata	&	56606.83400	&	4720	&	$B$	&	Image Fast	\\
\nodata	&	56606.88866	&	4720	&	$B$	&	Image Fast	\\
\nodata	&	56606.94403	&	4656	&	$B$	&	Image Fast	\\
\nodata	&	56606.99794	&	4720	&	$B$	&	Image Fast	\\
\nodata	&	56607.05259	&	4720	&	$B$	&	Image Fast	\\
\nodata	&	56607.10725	&	4720	&	$B$	&	Image Fast	\\
\nodata	&	56607.16190	&	4720	&	$B$	&	Image Fast	\\
\nodata	&	56607.21727	&	4656	&	$B$	&	Image Fast	\\
\nodata	&	56607.27118	&	4720	&	$B$	&	Image Fast	\\
\nodata	&	56607.32583	&	4720	&	$B$	&	Image Fast	\\
\nodata	&	56607.38049	&	4720	&	$B$	&	Image Fast	\\
\nodata	&	56607.43514	&	4720	&	$B$	&	Image Fast	\\
\nodata	&	56607.49051	&	4656	&	$B$	&	Image Fast	\\
\nodata	&	56607.54442	&	4720	&	$B$	&	Image Fast	\\
\nodata	&	56607.59907	&	4720	&	$B$	&	Image Fast	\\
\nodata	&	56607.65373	&	6520	&	$B$	&	Image Fast	\\
\nodata	&	56607.72921	&	4720	&	$B$	&	Image Fast	\\
\nodata	&	56607.78458	&	4656	&	$B$	&	Image Fast	\\
\nodata	&	56607.83850	&	6520	&	$B$	&	Image Fast	\\
\nodata	&	56607.91398	&	4720	&	$B$	&	Image Fast	\\
\nodata	&	56607.96863	&	4720	&	$B$	&	Image Fast	\\
\nodata	&	56608.02329	&	4720	&	$B$	&	Image Fast	\\
\nodata	&	56608.07866	&	2592	&	$B$	&	Image Fast	\\
\nodata	&	56608.10868	&	2656	&	$B$	&	Image Fast	\\
\nodata	&	56608.13944	&	2656	&	$B$	&	Image Fast	\\
\nodata	&	56608.17021	&	2656	&	$B$	&	Image Fast	\\
\nodata	&	56608.20097	&	2656	&	$B$	&	Image Fast	\\
\hline
\textit{Swift}/UVOT	&	56606.83056	&	1158	&	$UVW1$	&	Fast	\\
\nodata	&	56607.11389	&	2307	&	$UVW1$	&	Fast	\\
\nodata	&	56608.04375	&	1138	&	$UVW1$	&	Fast	\\
\nodata	&	56609.84792	&	1096	&	$UVW1$	&	Fast	\\
\hline
VLT/X-Shooter	&	56608.31458	&	3480	&	3000--25000\,{\AA}	&	Spectroscopy	\\
\hline
VLA	&	56606.68070	&	7200	&	4.5--5.5\,GHz	&	Imaging	\\
\nodata	&	56606.68070	&	7200	&	6.5--7.5\,GHz	&	Imaging	\\
\nodata	&	56607.57380	&	3600	&	2.0--4.0\,GHz	&	Imaging	\\
\hline
LOFAR	&	56607.18198	&     17818.8	&       0.139--0.162\,GHz       &	Imaging	\\
\hline
e-EVN	&	56609.08333	& 28800	        &	4.93--5.05\,GHz	&	Imaging	\\
\hline
Lovell	&	56607.06111	&	41640	&	1.3--1.7\,GHz	&	Timing	\\
\nodata	&	56608.05833	&	15360	&	1.3--1.7\,GHz	&	TIming	\\
\hline
WSRT	&	56607.02581	&	7190	&	2.20--2.34\,GHz	&	Timing	\\
\nodata	&	56607.11135	&	3579	&	0.31--0.38\,GHz	&	Timing	\\
\nodata	&	56607.15509	&	3580	&	2.20--2.34\,GHz	&	Timing	\\
\nodata	&	56607.19885	&	2380	&	0.31--0.38\,GHz	&	Timing	\\
\nodata	&	56608.02303	&	7190	&	2.20--2.34\,GHz	&	Timing	\\
\nodata	&	56608.10857	&	3579	&	0.31--0.38\,GHz	&	Timing	\\
\nodata	&	56608.15231	&	3580	&	2.20--2.34\,GHz	&	Timing	\\
\nodata	&	56608.19606	&	3580	&	4.83--4.97\,GHz	&	Timing	\\
\nodata	&	56608.23981	&	3580	&	2.20--2.34\,GHz	&	Timing	
\enddata
\label{obslog1}							 
\end{deluxetable*}

\begin{deluxetable*}{lcrcc}
\tabletypesize{\footnotesize}
\tablewidth{0pt}
\tablecaption{Log of Observations of PSR J1023+0038 during 2014 June 10--11.}
\tablehead{
\colhead{Telescope/} &
\colhead{Start time} & \colhead{Duration} & \colhead{Band/} & \colhead{Mode}\\
\colhead{Instrument} & \colhead{(MJD)} & \colhead{(s)} & \colhead{Filter} & \colhead{}}
\startdata
\textit{XMM-Newton}/EPIC pn	&	56818.18118	&	116726	&	0.3--10 keV	&	Fast timing	\\
\textit{XMM-Newton}/EPIC MOS1	&	56818.15704	&	13962	&	0.3--10	keV &	Small Window	\\
\nodata	& 	                        56818.33610	&	103225	&	0.3--10	keV &	Small Window	\\
\nodata	& 	                        56819.57051	&	2913	&	0.3--10	keV &	Small Window	\\
\textit{XMM-Newton}/EPIC MOS2	&	56818.15756	&	118502	&	0.3--10	keV &	Small Window	\\
\nodata	&	56819.57433	&	7301	&	0.3--10	keV &	Small Window	\\

\hline
\textit{Swift}/XRT	&	56819.16666	&	2763	&	0.3--10	keV &	Fast	\\
\hline
\textit{XMM-Newton}/OM	&	56818.16030	&	3456	&	B	&	Image Fast	\\
\nodata	&	56818.20032	&	3520	&	$B$	&	Image Fast	\\
\nodata	&	56818.24109	&	3520	&	$B$	&	Image Fast	\\
\nodata	&	56818.28185	&	3520	&	$B$	&	Image Fast	\\
\nodata	&	56818.32262	&	3520	&	$B$	&	Image Fast	\\
\nodata	&	56818.36410	&	5256	&	$B$	&	Image Fast	\\
\nodata	&	56818.42495	&	3520	&	$B$	&	Image Fast	\\
\nodata	&	56818.46572	&	3520	&	$B$	&	Image Fast	\\
\nodata	&	56818.50648	&	3520	&	$B$	&	Image Fast	\\
\nodata	&	56818.54725	&	3520	&	$B$	&	Image Fast	\\
\nodata	&	56818.58873	&	3456	&	$B$	&	Image Fast	\\
\nodata	&	56818.62875	&	3520	&	$B$	&	Image Fast	\\
\nodata	&	56818.66951	&	3520	&	$B$	&	Image Fast	\\
\nodata	&	56818.71028	&	3520	&	$B$	&	Image Fast	\\
\nodata	&	56818.79271	&	3520	&	$B$	&	Image Fast	\\
\nodata	&	56818.79252	&	3456	&	$B$	&	Image Fast	\\
\nodata	&	56818.83255	&	3520	&	$B$	&	Image Fast	\\
\nodata	&	56818.87331	&	3520	&	$B$	&	Image Fast	\\
\nodata	&	56818.91407	&	3520	&	$B$	&	Image Fast	\\
\nodata	&	56818.95484	&	3520	&	$B$	&	Image Fast	\\
\nodata	&	56818.99632	&	3456	&	$B$	&	Image Fast	\\
\nodata	&	56819.03634	&	3520	&	$B$	&	Image Fast	\\
\nodata	&	56819.07711	&	5320	&	$B$	&	Image Fast	\\
\nodata	&	56819.13870	&	3520	&	$B$	&	Image Fast	\\
\nodata	&	56819.17947	&	3520	&	$B$	&	Image Fast	\\
\nodata	&	56819.22095	&	3456	&	$B$	&	Image Fast	\\
\nodata	&	56819.26097	&	3520	&	$B$	&	Image Fast	\\
\nodata	&	56819.30174	&	3520	&	$B$	&	Image Fast	\\
\nodata	&	56819.34250	&	3520	&	$B$	&	Image Fast	\\
\nodata	&	56819.38326	&	3520	&	$B$	&	Image Fast	\\
\nodata	&	56819.42475	&	3456	&	$B$	&	Image Fast	\\
\nodata	&	56819.46477	&	3520	&	$B$	&	Image Fast	\\
\nodata	&	56819.50553	&	3520	&	$B$	&	Image Fast	\\
\nodata	&	56819.54630	&	3520	&	$B$	&	Image Fast	\\
\nodata	&	56819.58706	&	3520	&	$B$	&	Image Fast	\\
\hline
\textit{Swift}/UVOT	&	56819.16666	&	2761	&	$UVW2$	&	Fast	\\
\hline
WSRT	&	56819.52118	&	10800	&	1.30--1.46\,GHz	&	Timing	\\
\nodata	&	56819.64826	&	3600	&	2.20--2.34\,GHz	&	Timing	\\
\nodata	&	56819.69201	&	3600	&	0.31--0.38\,GHz	&	Timing
\enddata
\label{obslog2}							 
\end{deluxetable*}

\section{OBSERVATIONS AND DATA REDUCTION}
Tables 1 and 2 summarize all X-ray, UV/optical, and radio observations
presented in this paper, while Figure 1 shows a visual timeline of
this multi-wavelength observing campaign, arranged around two
 \textit{XMM-Newton} observations conducted in 2013 November
and 2014 June.

\subsection{X-ray Observations}

\subsubsection{\textit{XMM-Newton} EPIC}
PSR J1023+0038 was observed with \textit{XMM-Newton} starting on 2013
November 10 in a 134-ks exposure (ObsID 0720030101). It was revisited
on 2014 June 10 for 115 ks (ObsID 0742610101). For both observations,
the European Photon Imaging Camera (EPIC) pn instrument
\citep{struder01} was configured for fast timing mode, which permits a
30 $\mu$s time resolution at the cost of one imaging dimension. To
minimize the detrimental effect of event pile up, both EPIC MOS cameras
\citep{turner01} were used in small window mode with 0.3 s time
resolution. For all three detectors, the thin optical blocking filter
was in place.

The data reduction and extraction of the EPIC data were carried out
using the Science Analysis Software (SAS\footnote{The
  \textit{XMM-Newton} SAS is developed and maintained by the Science
  Operations Centre at the European Space Astronomy Centre and the
  Survey Science Centre at the University of Leicester.}) version {\tt
  xmmsas\_20130501\_1901-13.0.0}. The observations were filtered using
the recommended flag and pattern values. The analysis was
  restricted to the 0.3--10 keV range, over which the performance of
  the three detectors is well understood. For the X-ray variability
analysis, background-subtracted and exposure-corrected light curves
were obtained with the {\tt epiclccor} tool in SAS. Owing to the
bright nature of PSR J1023+0038 in its current state, when
constructing binned light curves it was not necessary to remove time
intervals of high backround flaring. On the other hand, to
maximize the sensitivity to pulsations, for the fast timing photon
lists, we determined time ranges corresponding to soft proton flares
by thresholding a 10-s binned light curve extracted from an off-source
region; photons arriving during these time ranges were not used in the
pulsed flux and profile analyses.

The pn fast timing mode data were extracted using a region of width
6.5 pixels in the imaging (RAWX) direction centered on row 37. This
translates to an angular size of 27$''$, which encircles $\sim$87\% of
the energy from the point source at 1.5 keV. The MOS1/2 source events
were obtained from circular regions of radius 36$''$ (limited by the
size of the small imaging window), which enclose $\sim$88\% of the
total point source energy at $\sim$1.5\,keV. Due to the occasional
instances of relatively high source count rates ($\sim$$4-5$\,counts
s$^{-1}$), the MOS1/2 instruments are susceptible to photon pile-up
even in small window mode. Pile-up occurs when two or more events
occur during a single read-out interval and are registered as a single
event with energy approximately equal to the sum of the individual
event energies \citep{Davis01}. This can result in artificial
hardening of the intrinsic source spectrum and loss of source
events. To diagnose the impact of pile-up, we used the SAS tool {\tt
  epatplot}. Based on this, we determined that pile-up is negligible.

For the variability and pulsation analyses, the photon arrival times
were translated to the solar system barycenter using the DE405 solar
system ephemeris and the best known astrometric position of the pulsar
from \citet{Del12}.

\subsubsection{\textit{XMM-Newton} RGS}
We extracted and processed the Reflection Grating Spectrometer (RGS)
data using the recommended SAS analysis procedures. The extracted
grating spectra show no obvious emission or absorption features. The
combination of lower count rate compared to the EPIC data and the
elevated background of the dispersed spectrum results in a low
signal-to-noise ratio. For this reason we do not make use of the RGS
data in our analysis.

\subsubsection{\textit{Swift} XRT}
\textit{Swift} has been used to regularly monitor the long-term
behavior of PSR J1023+0038 since 2013 October
\citep{Pat14,Tak14,Cot14}. It observed the binary on 2013 November
10--13 November 2013 and 2014 June 11 coincident with the
\textit{XMM-Newton} exposures.  The XRT was operated in
photon-counting mode (enabling 2.5-s time resolution) in all
exposures.  Due to the significantly lower sensitivity of the XRT
relative to the \textit{XMM-Newton} EPIC cameras, we do not use these
data in the analysis below.  We only note that, as expected, the
observed count rates are in full agreement with those observed with
\textit{XMM-Newton}.

\subsection{Optical/UV Photometry}

\subsubsection{\textit{XMM-Newton} OM}
The \textit{XMM-Newton} Optical Monitor \citep[OM;][]{Mason01} was used
with the $B$ filter, which has a band pass between 3800 and 5000\,{\AA}
centered on 4392\,{\AA}, and was employed in ``Image Fast'' mode to
provide high time resolution and photon counting capabilities. The
data were obtained in 30 exposures, typically of 4.7\,ks each, in the
2013 November observations and 35 exposures in the 2014 July
observation, typically of 3.5\,ks each (see Tables 1 and 2).  The
background-subtracted photometric $B$ filter data from the OM were
extracted using the SAS {\tt omfchain} pipeline script using the
default set of parameters. Time bins of width 10\,s were chosen in order to
examine the rapid variability in the system.

\subsubsection{\textit{Swift} UVOT}
We use all data collected with the \textit{Swift} Ultra-Violet/Optical
Telescope (UVOT) over the course of 2013 November 10--13 November and 2014 June
11. During the former, the UVOT was operated with the $UVW1$ filter,
which has a central wavelength of 2600\,{\AA} and a bandpass between
between 2200 and 4000\,{\AA}, while during the latter the $UVW2$ filter,
which has a central wavelength of 2246\,{\AA} and a bandpass between
1800 and 2600\,{\AA}, was in place.

We reduced the data using the UVOT pipeline in {\tt FTOOLS} and
applying standard event screening criteria.  We extracted source
events for each observation using circular regions with radius 5$''$
and by using the best source position available (Deller et
al.~2012). The resulting time series were barycentered to allow an
investigation of correlated behavior with the contemporaneous
\textit{XMM-Newton} light curves.

\begin{deluxetable}{ccccc}
\tabletypesize{\small}
\tablewidth{0pt}
\tablecaption{Log of MDM Observatory Time Series}
\tablehead{
\colhead{Telescope} & \colhead{Date (UT)} & \colhead{Time (UT)} &
\colhead{Filter} & \colhead{$N_{\rm exp}^a$}
}
\startdata
  1.3-m  &  2013 Dec 26  & 07:44--13:32  &  $B$     &  1598  \\
  1.3-m  &  2013 Dec 27  & 07:41--13:32  &  $B$     &  1606  \\
  1.3-m  &  2013 Dec 28  & 08:15--13:30  &  $B$     &  1445  \\
  1.3-m  &  2013 Dec 29  & 10:18--13:39  &  $BG38$  &   922  \\
  1.3-m  &  2014 Jan 3   & 08:33--13:36  &  $B$     &  1384  \\
  2.4-m  &  2014 Mar 22  & 06:52--09:41  &  $B$     &   670  \\
  2.4-m  &  2014 May 30  & 03:51--05:15  &  $B$     &   335  
\enddata
\tablenotetext{a}{All exposures were 10~s.}
\label{tab:log}
\end{deluxetable}

\subsubsection{MDM}
In this paper, we also present time-series optical photometry obtained
on seven nights during the 2013--2014 observing season using the MDM
Observatory's 1.3-m McGraw-Hill telescope or 2.4-m Hiltner telescope
on Kitt Peak (see Table 3 for a summary).  In all cases the detector
was the thinned, backside illuminated SITe CCD ``Templeton''. It has
$1024\times1024$ pixels, with a scale of $0.\!^{\prime\prime}509$
pixel$^{-1}$ on the 1.3-m, and $0.\!^{\prime\prime}275$ pixel$^{-1}$
on the 2.4-m.  In order to reduce the readout time, the CCD was
windowed down and the pixels were binned $2\times2$.  All exposures
were 10~s, with a read/prep time of 3\,s, resulting in a 13-s cadence.
All but one of the light curves used
the $B$ filter.  On 2013 December 29 we used a broadband $BG38$ filter.
Differential photometry was performed with respect to a field star
calibrated by \citet{Thor05} that has $B=15.51$ and $V=14.86$, the
latter used to approximate a magnitude in the $BG38$ filter.  A
heliocentric correction was applied to the observing times.

%
%
\begin{figure*}[!t]
\begin{center}
\includegraphics[width=0.95\textwidth]{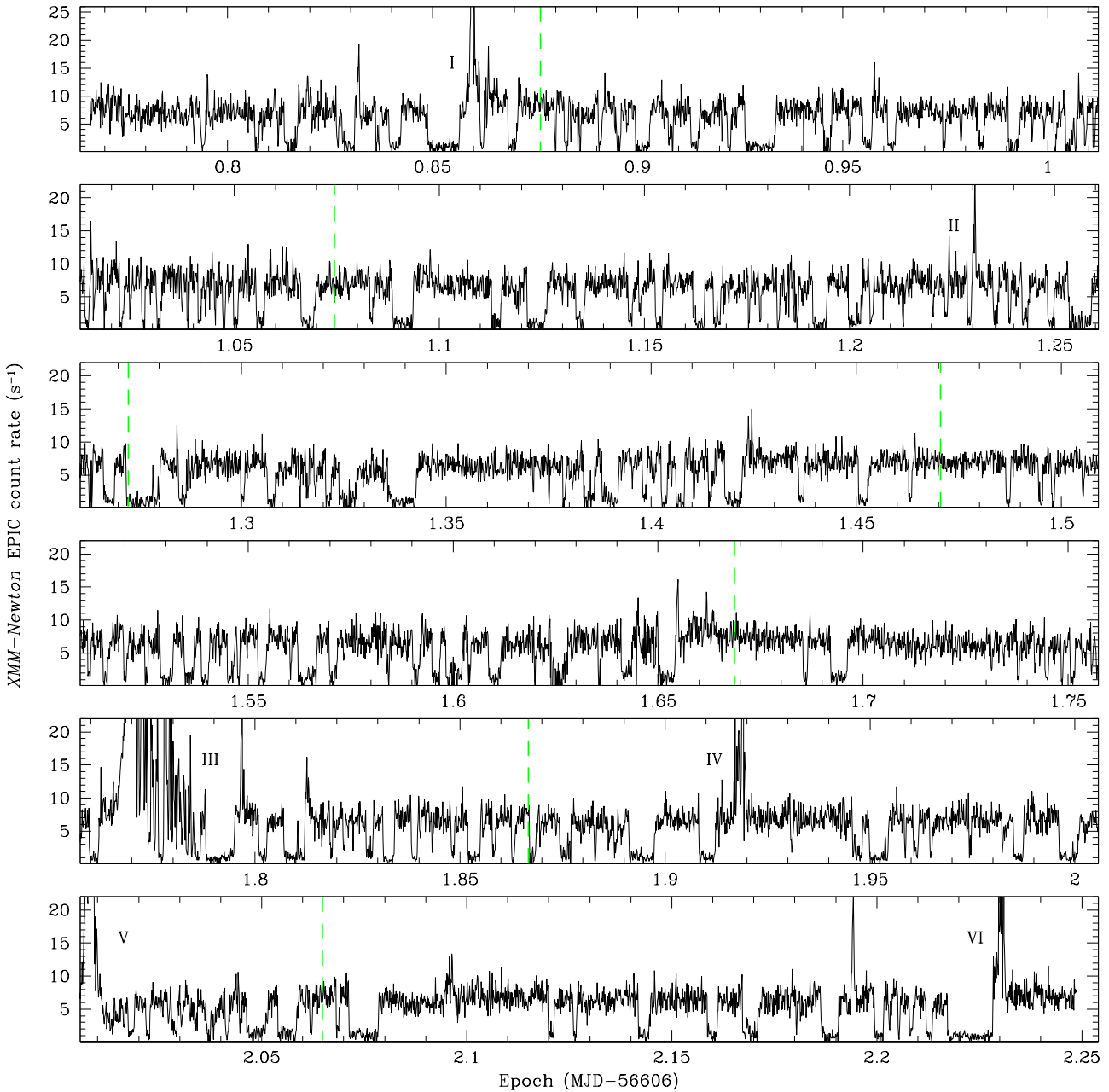}
\end{center}
\caption{Co-added, background-subtracted and exposure-corrected
  \textit{XMM-Newton} EPIC light curve of PSR J1023+0038 in the 0.3--10
  keV band from ObsID 0720030101, acquired during 2013 November
  10--12. The vertical dotted lines mark the times of orbital phase
  $\phi_b=0.25$ of the binary. The Roman numerals mark the six
  prominent flares.}
\end{figure*} 

%
%
\begin{figure*}[!t]
\begin{center}
\includegraphics[width=0.95\textwidth]{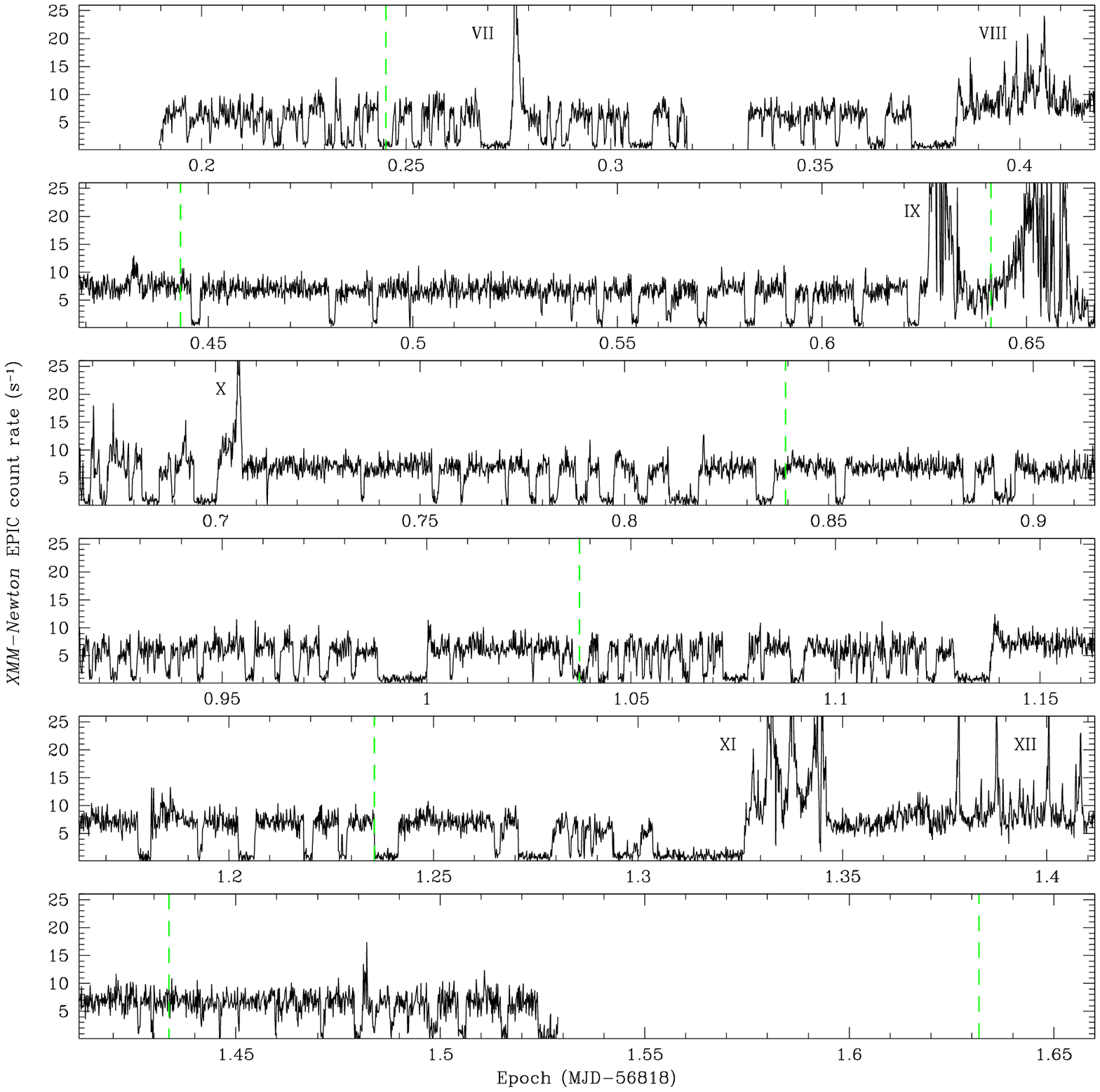}
\end{center}
\caption{Same as Figure 2 but for ObsID 0742610101, acquired during 2014
  June 10--11. }
\end{figure*}

\subsection{Optical Spectroscopy}

PSR\,J1023+0038 was observed with the X-SHOOTER instrument at the
European Southern Observatory (ESO) Very Large Telescope (VLT) in Paranal,
Chile. X-SHOOTER is a medium resolution ($R=4000$ to 7000)
spectrograph covering a wavelength range from 3000\,\AA\ to
2.5\,$\mu$m. Two exposures were obtained on 2013 November 12 between
07:55UT and 08:30UT, shortly after the end of the 2013 November
\textit{XMM-Newton} observation. The exposure times were 600\,s in the
UVB arm (3000--5595\,\AA), 628.7\,s in the VIS arm (5595\,\AA\ to
1.024\,$\mu$m) and $2\times300$\,s in the NIR arm
(1.024--2.48\,$\mu$m). The width of the slit was $1\farcs0$ in the UVB
arm and $0\farcs9$ in the VIS and NIR arms. The observing conditions
were good with $1\farcs1$ seeing. The obtained spectra were reduced
and calibrated with standard ESO software tools (Reflex 2.6, X-Shooter
pipeline 2.5.2).

\subsection{Radio Timing Observations}

For part of both the November 2013 and June 2014 observations,
simultaneous radio data were acquired with either one, or both, of the
Westerbork Synthesis Radio Telescope (WSRT) and the Lovell Telescope (LT)
at Jodrell Bank Observatory (see Tables 1 and 2 for details). The WSRT
observations were made at central frequencies of 350, 1380, 2273,
and 4901\,MHz and the LT observations were made at 1532\,MHz. A
detailed description of the observing systems can be found in
\citet{Stap14}.

\subsection{Radio Imaging Observations}

PSR J1023+0038 was targeted by the Karl G.~Jansky Very Large Array
(VLA) in B configuration for two hours on 2013 November 10, covering
4.5$-$5.5 and 6.5$-$7.5 GHz, and for one hour on 2013 November 11,
covering 2.0$-$4.0 GHz.

The system was observed with the very long baseline interferometry
technique (VLBI), with the European VLBI Network (EVN) in realtime
e-VLBI mode on 2013 November 13 between 2:00--10:00 UT. The following
telescopes of the e-EVN array participated: Jodrell Bank (MkII),
Effelsberg, Hartebeesthoek, Medicina, Noto, Onsala (25m), Shanghai
(25m), Toru\'n, Yebes, and the phased-array WSRT.

PSR J1023+0038 was also observed by the Low Frequency Array
\citep[LOFAR;][]{van-haarlem13a} for 5 hours on 2013 November 11,
between 04:22--09:19 UT, with bandwidth spanning the frequency
range 138.5--161.7 MHz.  All Dutch stations were included, for a total
of 60 correlated elements.

A more detailed description of the calibration, data reduction and
analysis procedures of all radio imaging observations --- including a
much more extensive radio continuum monitoring campaign --- is provided
in \citet{Del14}.
 
%
%
\begin{figure*}[!t]
\begin{center}
\includegraphics[width=0.46\textwidth]{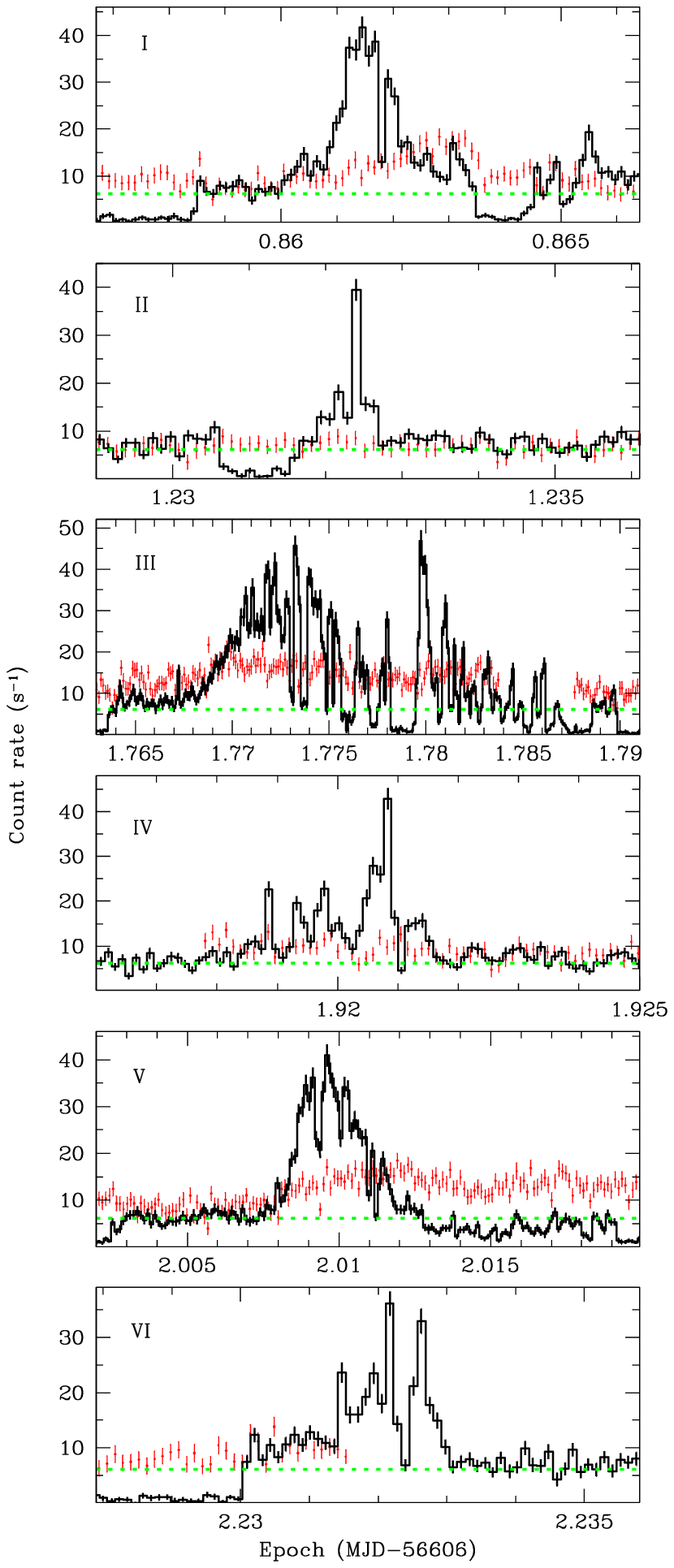}
\includegraphics[width=0.406\textwidth]{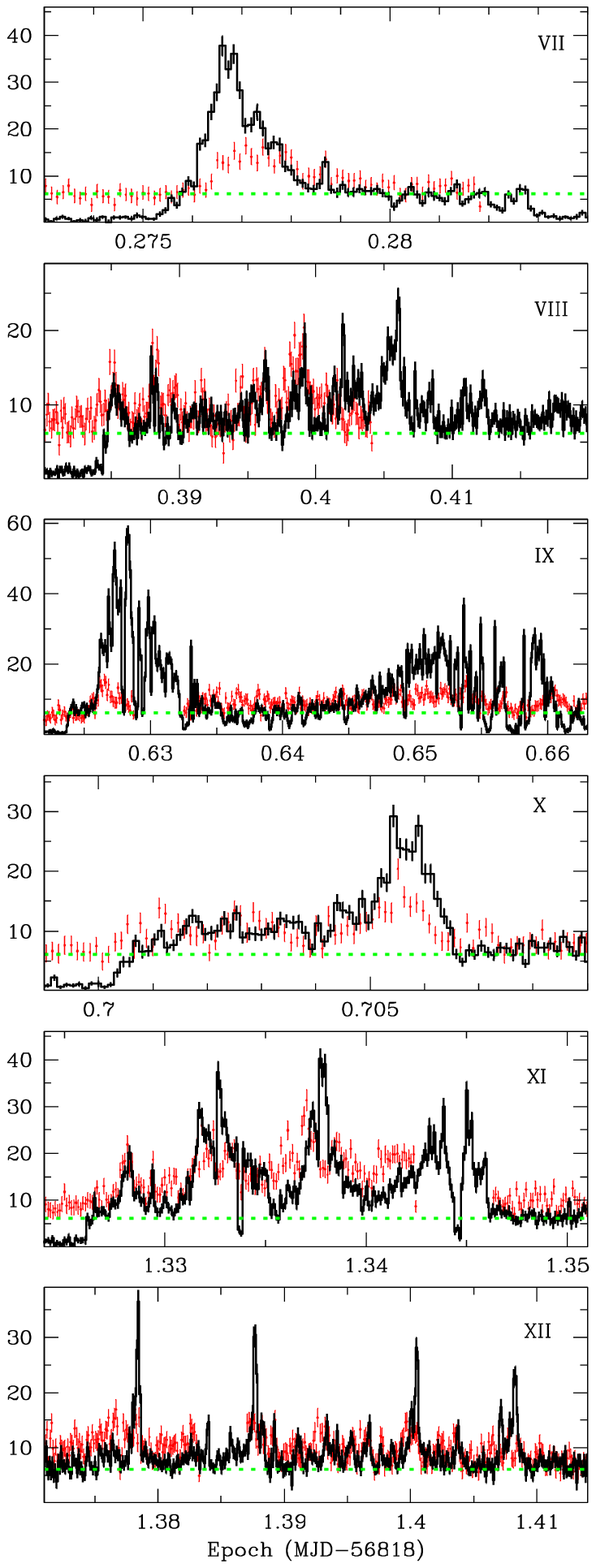}
\end{center}
\caption{A detailed view of the prominent flares observed with
  \textit{XMM-Newton} in 2013 November (left) and 2014 June
  (right). The Roman numerals correspond to those in Figures 2 and
  3. The red points show the photometric $B$ filter data from the
  \textit{XMM-Newton} OM in units of counts s$^{-1}$ binned at a 10 s
  resolution.  For reference, the dotted (green) horizontal line shows the
  mean X-ray flux level in the 0.3--10 keV range over the entire
  observation.}
\label{fig:flares}
\end{figure*}

\section{X-ray Variability}
Figures 2 and 3 show the total exposure-corrected,
background-subtracted \textit{XMM-Newton} EPIC X-ray light curves in
the 0.3--10 keV band, obtained by combining the data from the pn,
MOS1, and MOS2 during the periods when all three telescopes acquired
data simultaneously. The large-amplitude variability is obvious and is
reminiscent of the behavior seen in the \textit{Swift} XRT
\citep{Pat14} and \emph{NuSTAR} \citep{Ten14} observations of PSR
J1023+0038 from 2013 October. During both \textit{XMM-Newton}
obervations, most of the time is spent in the ``high''
mode\footnote{For the sake of clarity, we adopt the following
  nomenclature: we refer to the three distinct X-ray flux levels as
  ``modes'', while we refer to the long-term radio pulsar and LMXB
  intervals as ``states''.}, with a typical total EPIC count rate of
$\approx$7 counts s$^{-1}$ (0.3--10 keV).  The emission drops out
unpredictably to the ``low'' mode with count rate $\approx$1 counts
s$^{-1}$ with very rapid ingress and egress ($\sim$10--30 s).

Sporadic, intense X-ray flares (labeled by the Roman numerals in
Figures 2 and 3), reaching up to $\approx$60 counts s$^{-1}$,
occur on average every few hours, and exhibit diverse morphologies and
durations. Some flares last less than a minute (e.g. II), while others
(e.g. III, IX, and XI) last up to $\approx$45 minutes. The long-duration
flares exhibit a great deal of intricate structure. Specifically,
throughout flares III and XI, there are a number of rapid drops in
count rate that occasionally reach down to the levels of the low flux
mode. In both \textit{XMM-Newton} observations, the reccurence time
between X-ray flares is comparable ($\sim$20\,ks), with six prominent
flares (which we define as those with peak rates exceding 25
counts s$^{-1}$) in each observation.

It is interesting to note that the brighter flares are usually
  (but not always) preceded by a low mode instance (see, e.g., flares
  I, IV, V, VI, VII, VIII, X, and XI). This behavior could be an
  indication of a physical process that depends on the accumulation of
  a reservoir of energy (during the low mode) that is rapidly released
  (in the flare). However, due to the limited number of strong flares
  in the present data combined with the fairly frequent occurrence of
  low mode intervals, it is not possible to determine with certainty
  if this is merely a chance coincidence or if the flares and their
  corresponding pre-flare low mode are truly associated.  The absence
  of low modes immediately preceding some flares argues against a
  causal connection because it implies that the occurence of a low
  mode interval is not required to initiate a flare.

In the disk-free, radio pulsar state, PSR J1023+0038 exhibited
pronounced orbital-phase-dependent X-ray flux modulations
\citep{Arch10,Bog11}. Although in the light curves shown in Figures 2
and 3 there is no obvious periodic behavior (see \S4), in principle,
orbital modulation may still be present in one of the X-ray flux
modes.  To investigate this possibility, we separately extracted the
emission from the high, low, and flare modes and folded them at the
orbital period.  We do not find evidence for orbital modulation of the
X-ray brightness during any of the three modes.

\citet{Li14}  have reported evidence for periodicity
at 3130 s in the \textit{NuSTAR} data set of PSR J1023+0038. We have
examined both \textit{XMM-Newton} observations in search of a similar
signal but find no evidence for it. In addition, we are not able to
reproduce this result using the \textit{NuSTAR} data presented in
\citet{Ten14}.

\begin{figure}
\centering
\includegraphics[width=0.45\textwidth]{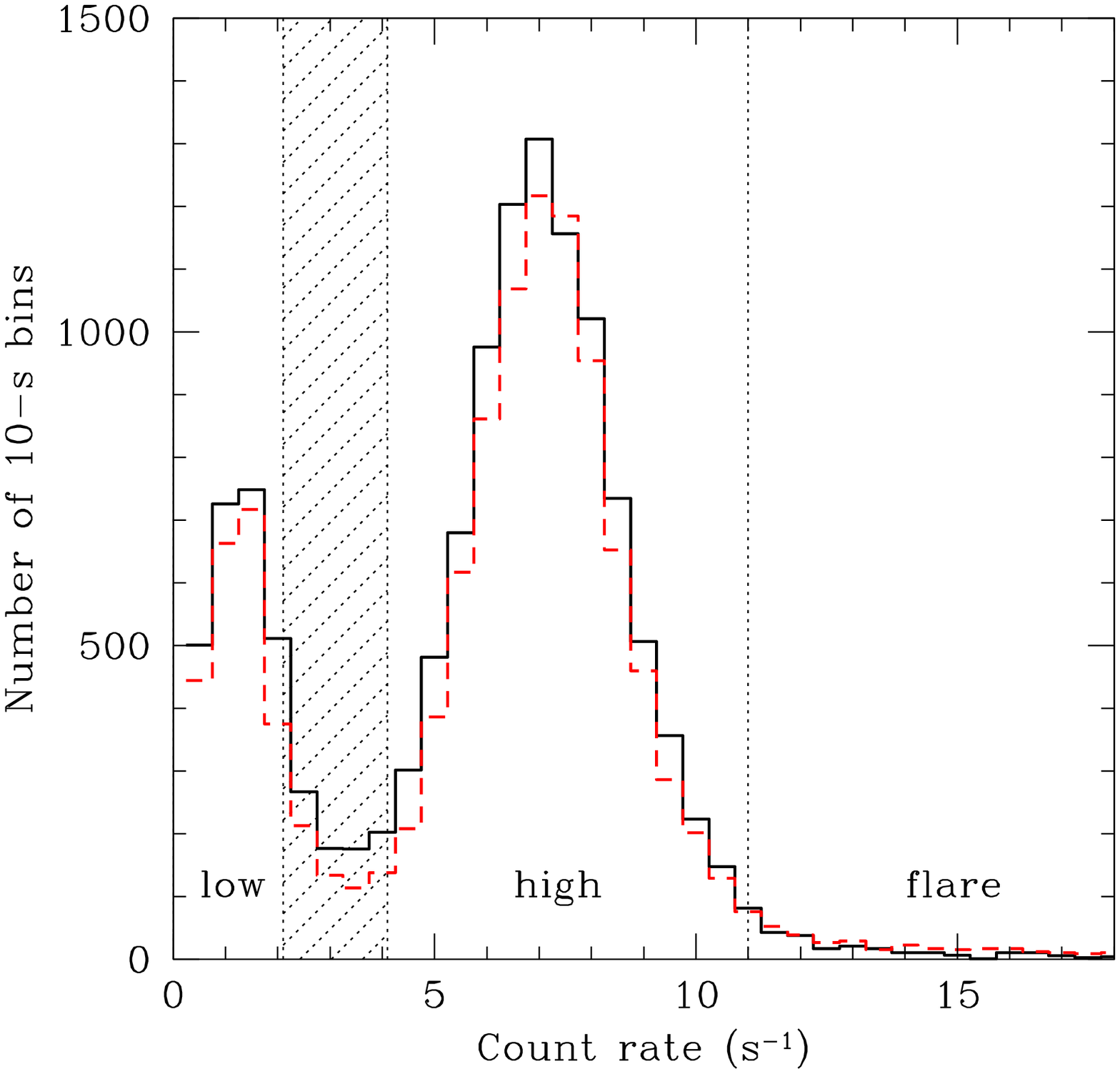}
\caption{Distribution of count rates obtained from the 10-s binned
  light curves from the 2013 November (solid black line) and 2014 June
  (dashed red line) \textit{XMM-Newton} data.  We define the region
  between $0$\,counts s$^{-1}$ and $2.1$\,counts s$^{-1}$ as the low
  mode, while the zone between $4.1$\,counts s$^{-1}$ and $11$\,counts
  $s^{-1}$ is the high mode. The hatched region denotes the ``grey
  area'' or transition region defined in the text.  }
\label{fig:xmm_poisson_smoothed_histogram}
\end{figure}

\begin{figure}
\centering \includegraphics[clip=true,trim=0.05in 0.0in 0.55in
  0.0in,width=0.45\textwidth]{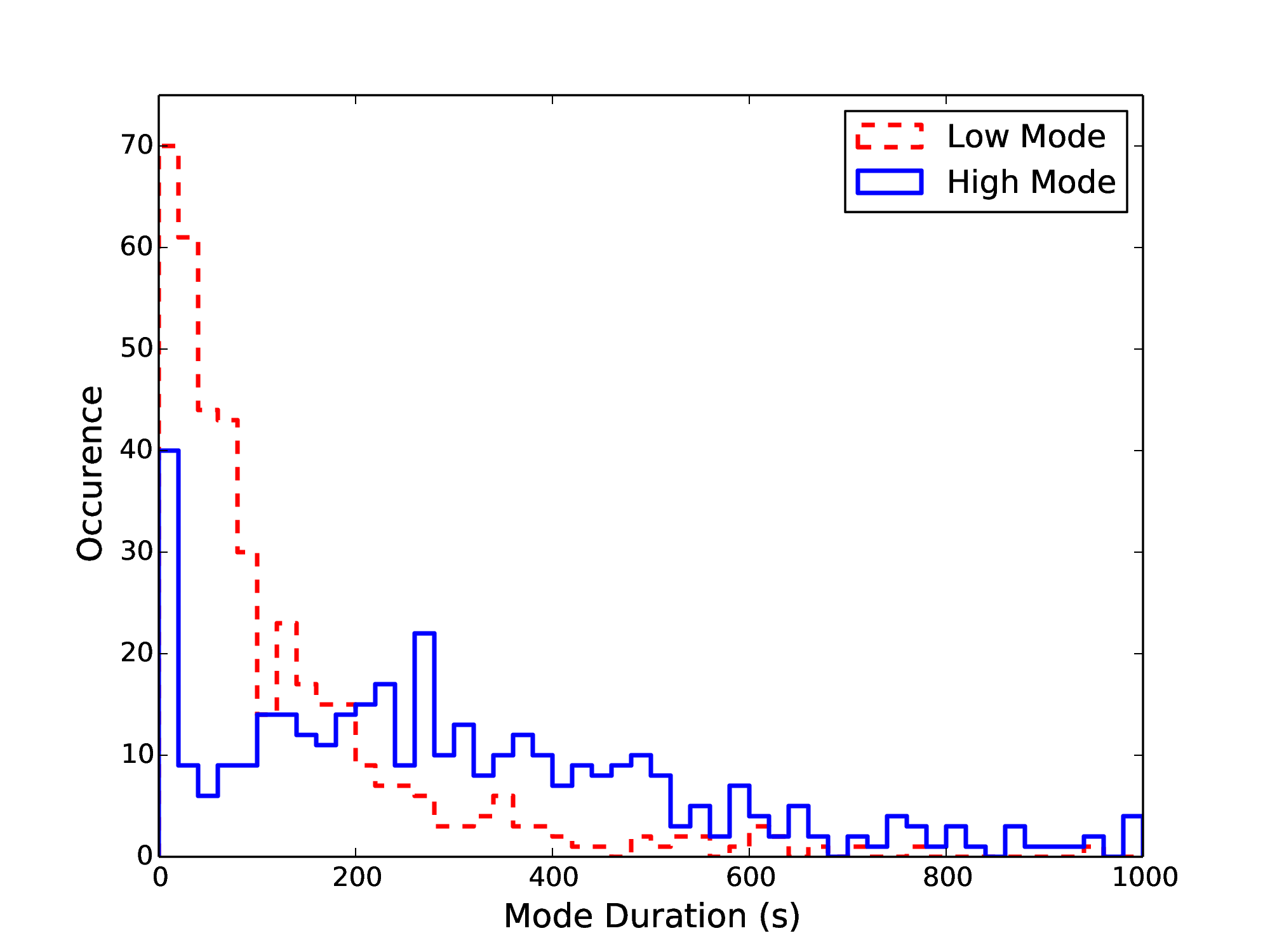}
\caption{Histogram of the durations of low (dashed red line) and
  high modes (solid blue line). The data from 2013 November and 2014 June
  are combined. While the longest high mode instance observed is
  $\approx 7000$\,s long, this histogram is truncated to durations
  $<$1000\,s to improve visibility.}
\label{fig:xmm_hi_state_lo_state_durations}
\end{figure}

\section{Statistical Analysis of the X-ray Variability}
We performed a statistical analysis of the low, high, and flare X-ray
flux modes to understand their temporal properties and identify any
correlations between them. We followed the same methodology as the
analysis performed on the \nustar\ data of PSR J1023+0038 in
\citet{Ten14}. We use the combined background-subtracted and exposure
corrected EPIC 0.3--10 keV data binned with 10-s bins shown in Figures
2 and 3. The 10-s bins were chosen to ensure sufficient photon
statistics in the low mode bins and in the mode transitions.

Figure~\ref{fig:xmm_poisson_smoothed_histogram} shows the distribution
of count rates throughout the 2013 November (solid black line) and
2014 June (dashed red line) observations.  The bi-modality in the
distribution is immediately apparent, with the low and high modes
clearly seen as peaks at $\approx$1\,counts s$^{-1}$ and
$\approx$7\,counts s$^{-1}$ respectively for both epochs. In both
instances, the distribution of count rates for the low and high mode
are consistent with what is expected from a Poisson distribution.  The
minimum of the distribution between the two states occurs at
approximately 3.1\,counts s$^{-1}$.

To cleanly differentiate between the two modes and the transitions
between them, we define a ``grey area'' (hatched region in
Figure~\ref{fig:xmm_poisson_smoothed_histogram}) ranging from 2.1\,counts
s$^{-1}$ to 4.1\,counts s$^{-1}$. These limits were chosen to be
symmetrical around the minimum of the distribution, though because of
the inherent count-rate errors in each light curve bin, the exact
values of these thresholds is not critical to the analysis presented
below\footnote{In \citet{Arch14}, slightly different criteria were
  used in defining the modes. Nevertheless, the two procedures result
  in negligible differences ($\lesssim$1\%).}. In particular, varying
the thresholds by $\pm$0.3 s$^{-1}$ results in a difference of less
than 1\% in the number of transitions.
Figure~\ref{fig:xmm_poisson_smoothed_histogram} shows the
  count rate levels used to define the low and high states. In the
analysis that follows, the flare mode intervals were removed from the
light curves by excising intervals with rates in excess of $11$ counts
s$^{-1}$.

We use a bi-stable comparator\footnote{In the field of electronics
  this is known as a Schmitt trigger.} to define the transitions
between the modes as follows. A low$\to$high transition is defined
when the count rate crosses from the ``low'' zone to the ``high'' zone
shown in Figure~\ref{fig:xmm_poisson_smoothed_histogram} and reverse
for the high$\to$low transition. If the light curve varies from the
low region to the grey area and returns back to the low region, the
light curve mode is maintained as low and a transition is not
registered. Similarly, a variation from the high zone to the grey area
and back to the high zone is maintained as a high mode.  Using this
comparator algorithm, we counted 237 low$\to$high and high$\to$low
transitions in the 2013 November data and 172 low-high transitions and
171 high-low transitions in the 2014 June data. We measured the
durations of the low and high modes as the number of light curve bins
between the end of the previous transition and the beginning of the
next transition. The duration of the transition itself was estimated
as the number of 10-s light curve bins spent in the grey area.

\begin{figure*}
\centering
\includegraphics[width=0.7\textwidth]{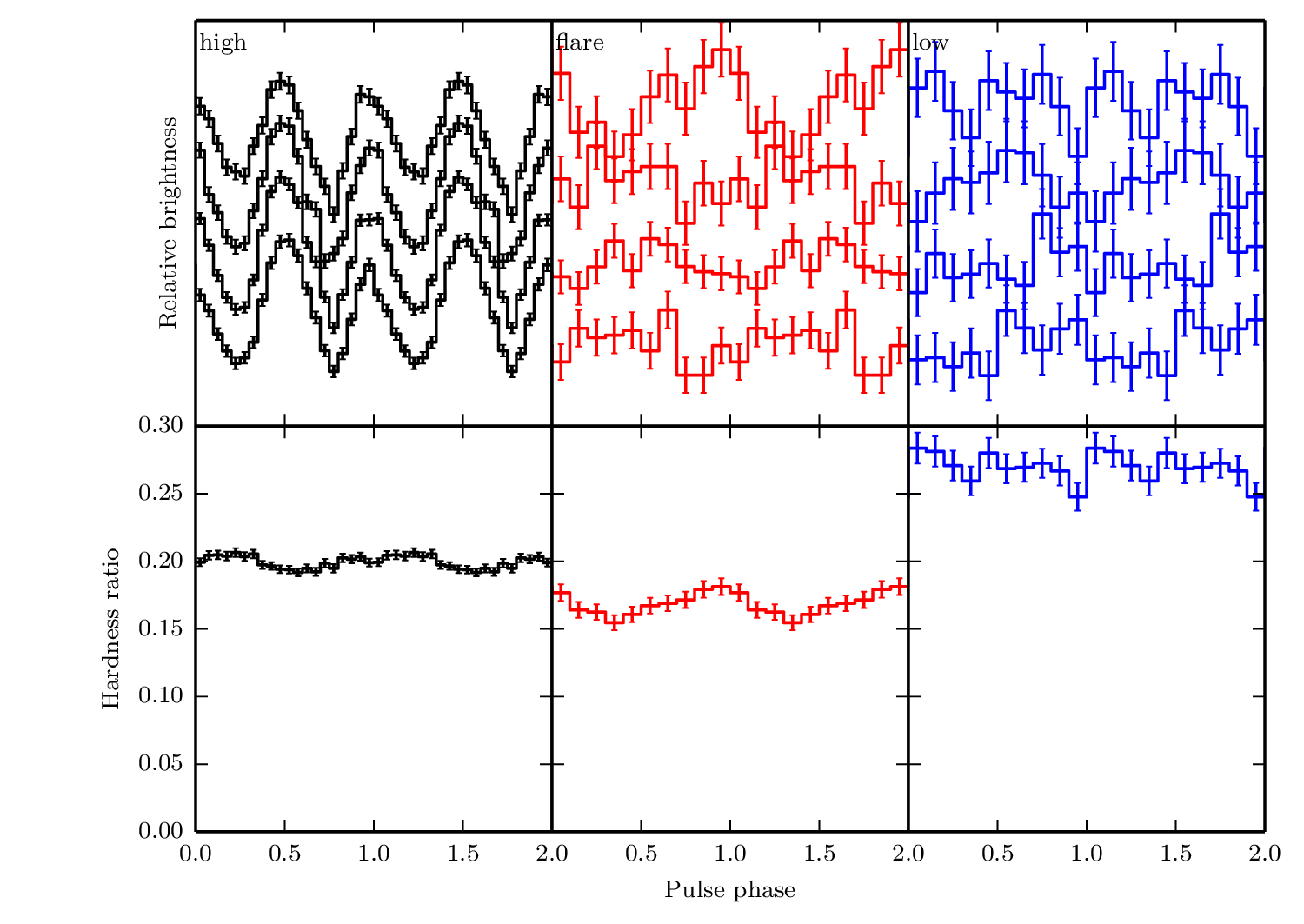}
\caption{Top panels: Normalized pulse profiles of PSR J1023+0038 in
  the 0.3--0.7, 0.7--1.5, 1.5--3, and 3--10 keV bands (from bottom to
  top, respectively) in the three X-ray flux modes: high (left), flare
  (middle), and low (right). Bottom panels: Hardness ratios of the
  three pulse profiles with the 1--10 keV counts divided by the counts
  in the 0.3-1 keV band.}
\label{fig:pulse_prof}
\end{figure*}

\begin{figure}
\centering
\includegraphics[width=0.42\textwidth]{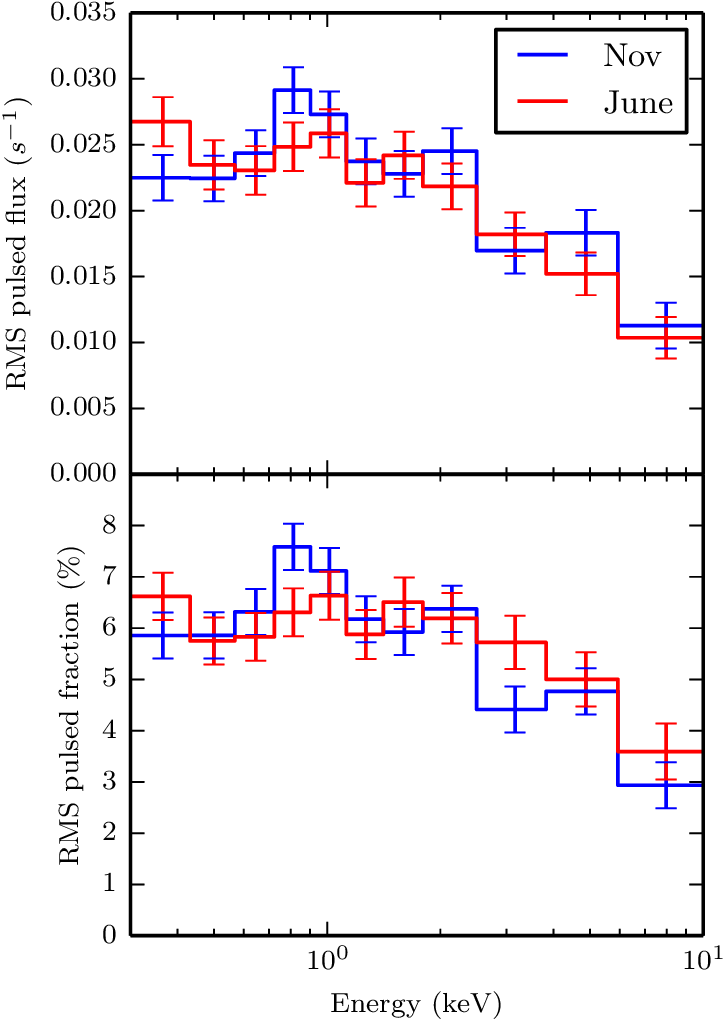}
\caption{Pulsed flux and fraction as a function of energy for both
  November and June observations. The flux curve (top), expressed in
  instrument counts per second, is dominated by instrumental
  sensitivity, but only the relatively low background distorts the
  shape of the pulsed fraction curve (bottom).}
\label{fig:pulse_energy}
\end{figure}

Figure~\ref{fig:xmm_hi_state_lo_state_durations} shows the
distribution of durations of the low (dashed red line) and high (blue
line) mode based on the combined 2013 November and 2014 June data
sets. The histogram is truncated to show only durations shorter than
1000\,s although the longest continuous high mode interval was 6910\,s
in 2014 June and the longest low mode lasted 1930\,s, also in 2014
June. We find that the low mode exists for $\approx$$22$\% of the
total observation time during 2013 November 10-12 and $\approx$$21$\%
during 2014 June 10-11 observations.  As shown in
Figure~\ref{fig:xmm_hi_state_lo_state_durations}, on average the high
mode tends to persist for significantly longer durations than the low
mode. The distribution of low mode durations can be very
  approximately modeled as a power law with index $-1.1$ and $-1.3$,
  for the 2013 November and 2014 June observations, respectively. The
  separations of the centroids of consecutive low modes follow a
  log-normal distribution with mean separations of $\sim$441 and
  $\sim$403 s and standard deviations of $0.93$ and $0.76$ for the two
  data sets, respectively.

We also looked for correlations or patterns between durations of
consecutive modes.  There is no evidence for correlated behavior
between the two modes, indicating a stochastic nature of the
underlying physical mechanism responsible for the switches between the
two. However, we do note that the modes tend to last longer during
2014 June as compared to the durations in 2013 November, which
suggests some long-term variation in behavior.

We investigated the distribution of the times of the low-high and
high-low transitions for 2013 November and 2014 June as well.  We find
that 57\% of the high$\to$low transitions in 2013 November and 54\% of
those in 2014 June lasted $<$$10\,$s. In contrast, only 40\% of the
low$\to$high transitions in 2013 November and 39\% of those in 2014 June
were $<$$10\,$s. Thus, based on this estimate, the high$\to$low transitions
may be faster than the low$\to$high transitions.

\begin{deluxetable}{lcrcc}
\tabletypesize{\footnotesize}
\tablewidth{0pt}
\tablecaption{Results of absorbed power-law fits of the \textit{XMM-Newton} spectra of PSR J1023+0038.}
\tablehead{
\colhead{ } & \colhead{$N_{\rm H}$} & \colhead{$\Gamma$} & \colhead{$L_X$\tablenotemark{b}} & \colhead{$\chi^2$/dof}  \\
\colhead{Mode} & \colhead{($10^{20}$ cm$^{-2}$)} & \colhead{ } & \colhead{($10^{33}$ erg s$^{-1}$)} & \colhead{}  }
\startdata
 \multicolumn{5}{c}{2013 November 10-12 (ObsID 0720030101)}\\
\hline
\textit{Flare}  & $2.7(9)$   & $1.65(4)$  & $10.8(2)$  &  $0.98/321$  \\
\textit{High}  &  $3.1(2)$ &  $1.71(1)$   & $3.17(2)$  & $1.23/902$ \\
\textit{Low}   &  $2.6(11)$ & $1.80(5)$    & $0.54(1)$  & $1.02/281$  \\
\hline
\textit{Flare}\tablenotemark{c}  & $3.1(2)$   & $1.66(2)$  & $10.9(2)$  &  $1.14/1506$  \\
\textit{High}                    &  \nodata &  $1.71(1)$   & $3.17(2)$  & \nodata \\
\textit{Low}                     &  \nodata & $1.82(3)$    & $ 0.54(1)$  & \nodata  \\
\hline
\multicolumn{5}{c}{2014 June 10-11 (ObsID 0742610101)}\\
\hline
\textit{Flare}  & $3.6(8)$   & $1.76(3)$   & $9.6(2)$  & $1.02/373$   \\
\textit{High}  &  $3.1(2)$ &  $1.75(1)$  & $3.06(2)$  & $1.31/870$ \\
\textit{Low}  &  $2.4(12)$ & $1.77(6)$   & $0.45(1)$ & $0.96/206$ \\
\hline
\textit{Flare}\tablenotemark{c}  & $3.1(2)$   & $1.74(2)$   & $9.6(1)$  & $1.18/1452$   \\
\textit{High}                    &  \nodata  &  $1.75(1)$  & $3.06(2)$  & \nodata \\
\textit{Low}                     &  \nodata   & $1.80(4)$   & $0.46(1)$ & \nodata
\enddata
\tablenotetext{a}{The numbers in parentheses show the 90\% confidence level uncertainties in the last digit of the quoted best-fit values.}
\tablenotetext{b}{Luminosity in the 0.3--10 keV band assuming the parallax distance of 1.37 kpc \citep{Del12}.}
\tablenotetext{c}{Fits performed jointly on the spectra of all three modes with a tied value of $N_{\rm H}$.}
\end{deluxetable}

\begin{deluxetable*}{lcrcccccc}
\tabletypesize{\footnotesize}
\tablewidth{0pt}
\tablecaption{Multi-component model fits of the high mode \textit{XMM-Newton} spectra of PSR J1023+0038.}
\tablehead{
\colhead{ } & \colhead{$N_{\rm H}$} & \colhead{$\Gamma$} &  \colhead{$T_{\rm eff}$} &  \colhead{$R_{\rm eff}$} & \colhead{$L^{\infty}/L_{\rm Edd}$} &  \colhead{$L_{X,PL}$\tablenotemark{b}}  & \colhead{$L_X$\tablenotemark{b}} & \colhead{$\chi^2$/dof}  \\
\colhead{Model\tablenotemark{a}} & \colhead{($10^{20}$\,cm$^{-2}$)} & \colhead{ } & \colhead{($10^6$\,K)} &  \colhead{(km)} & \colhead{ } & \colhead{($10^{33}$\,erg s$^{-1}$)} & \colhead{($10^{33}$\,erg s$^{-1}$)} & \colhead{}  }
\startdata
 \multicolumn{9}{c}{2013 November 10-12 (ObsID 0720030101)}\\
\hline
\textit{powerlaw + nsa}\tablenotemark{c}  &  $2.5(6)$ &  $1.59(3)$   & $1.49(21)$ & $2.9^{+2.6}_{-1.9}$ & \nodata &  $3.00(4)$ & $3.18(3)$  & $1.12/900$ \\
\textit{powerlaw + zamp}  &  $2.5(5)$ &  $1.60(3)$   & \nodata & \nodata &  $-4.6(2)$      &  $3.02(4)$ &   $3.18(3)$  & $1.12/900$ \\
\hline
\multicolumn{9}{c}{2014 June 10-11 (ObsID 0742610101)}\\
\hline
\textit{powerlaw + nsa}  &  $2.3(6)$ &  $1.60(3)$  & $1.45(18)$ & $3.3^{+2.7}_{-1.9}$ & \nodata & $2.85(4)$ & $3.06(3)$  & $1.15/868$ \\ 
\textit{powerlaw + zamp}  &  $2.2(5)$ &  $1.61(3)$  & \nodata & \nodata &  $-4.6(2)$ & $2.87(4)$ & $3.06(3)$  & $1.15/868$ 
\enddata
\tablenotetext{a}{All uncertainties quoted are at a 90\% confidence level.}
 \tablenotetext{b}{Luminosity in the 0.3--10 keV band assuming the parallax distance of 1.37 kpc \citep{Del12}.}
\tablenotetext{c}{For the \textit{nsa} model a neutron star with $M=1.4$ M$_{\odot}$ and $R=12$ km is assumed.}
\end{deluxetable*}

%
%
\begin{figure*}[!t]
\begin{center}
\includegraphics[width=0.4\textwidth]{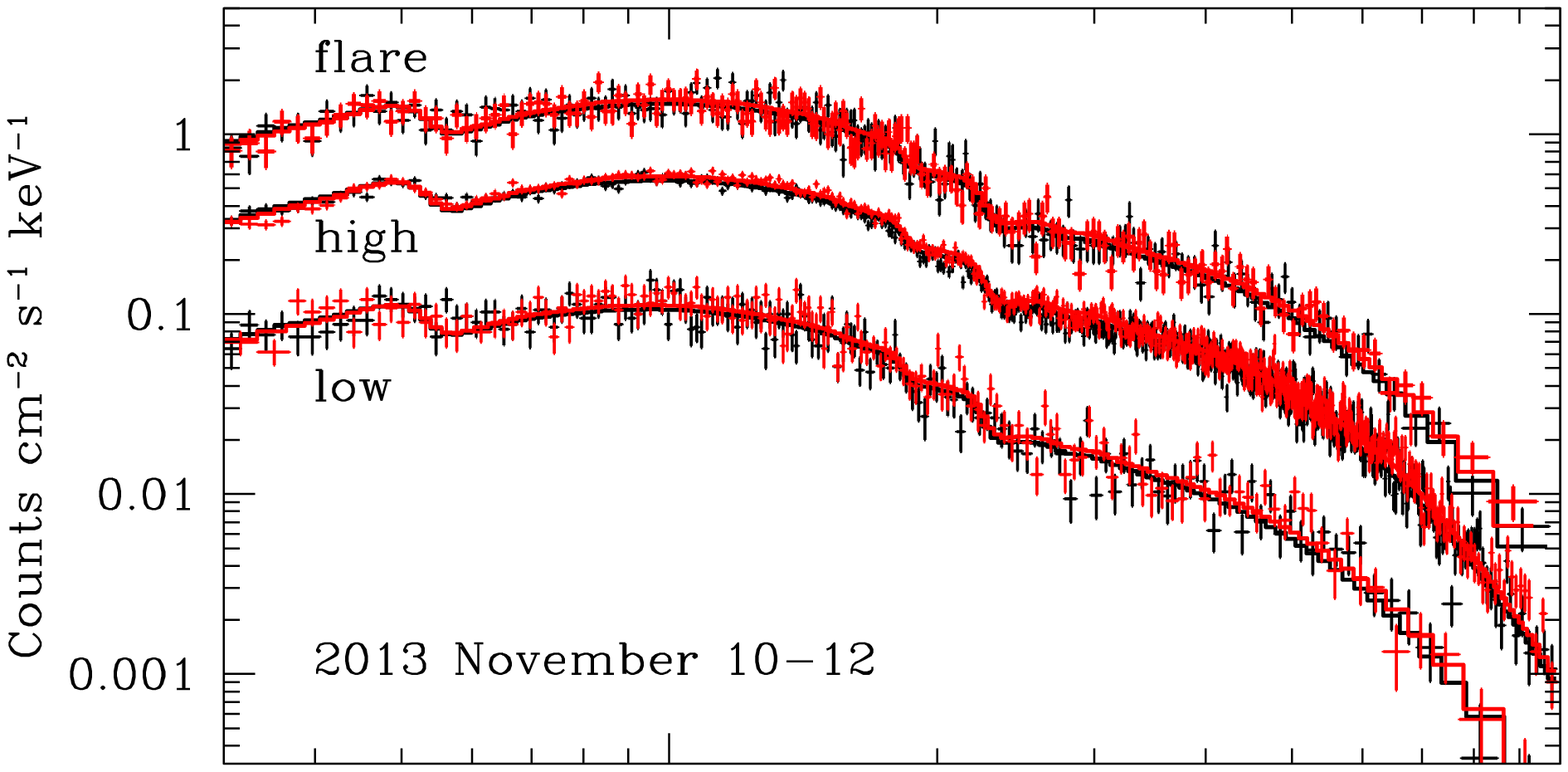}
\includegraphics[width=0.3545\textwidth]{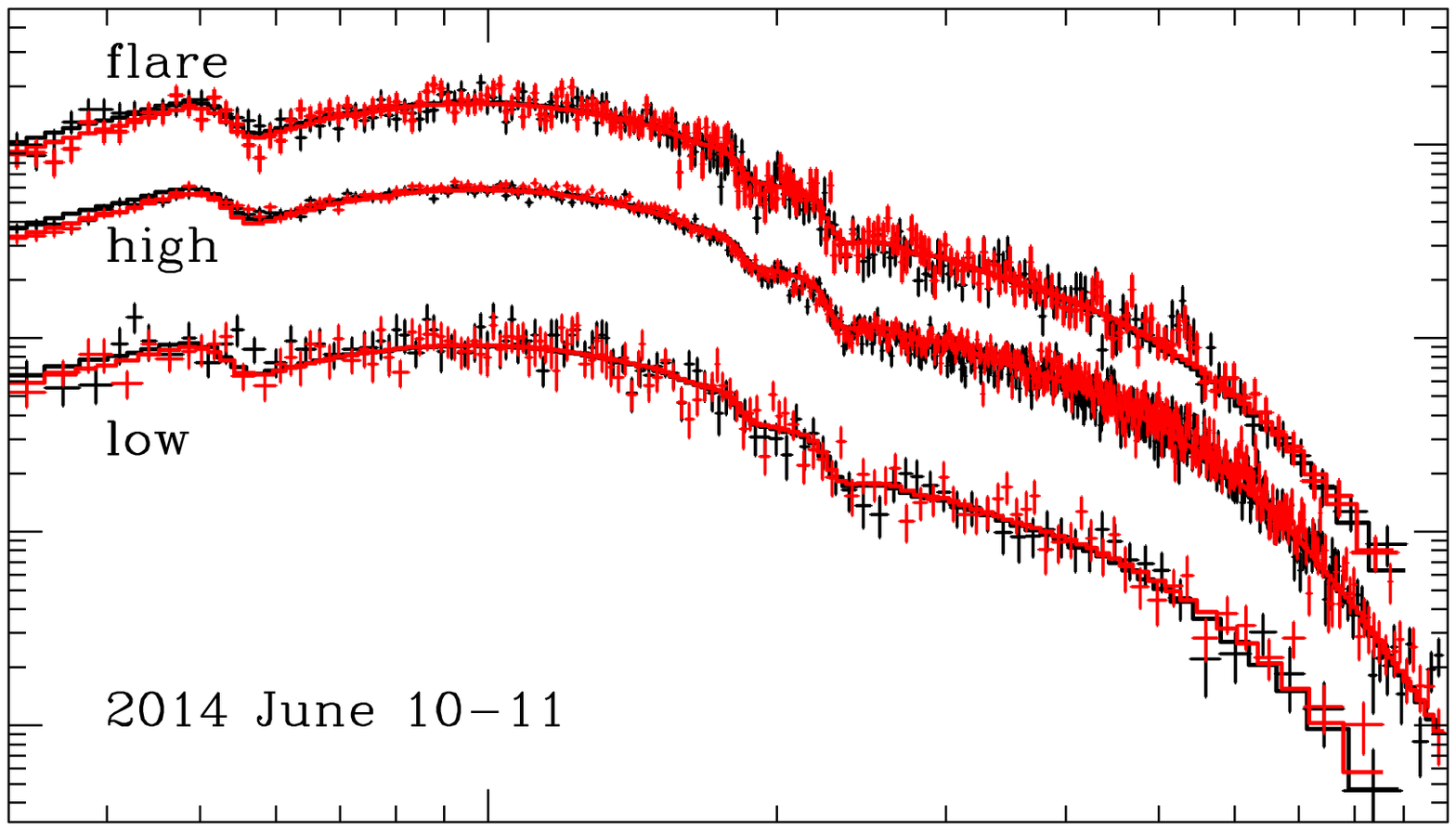}
\includegraphics[width=0.4\textwidth]{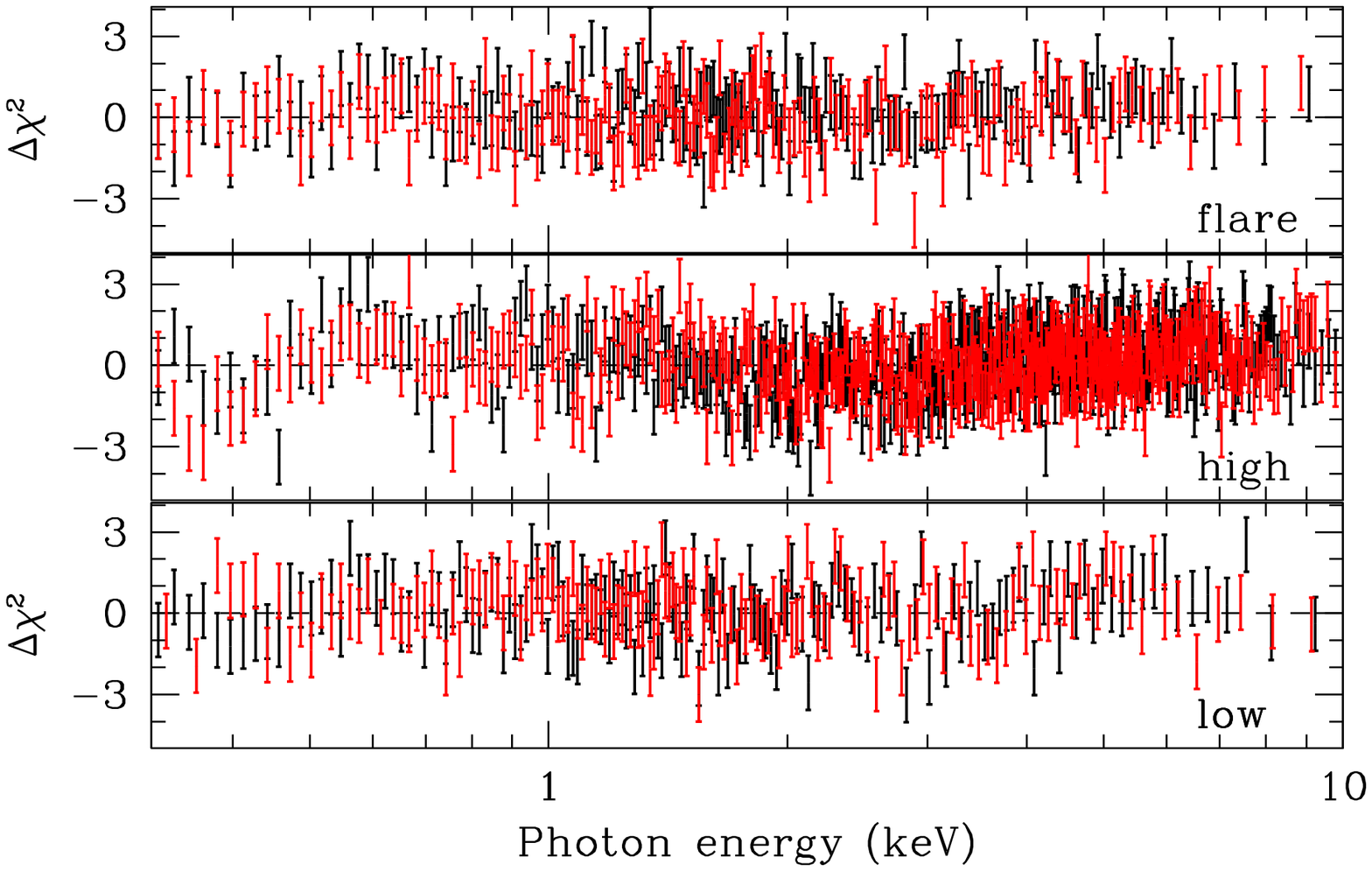}
\includegraphics[width=0.3545\textwidth]{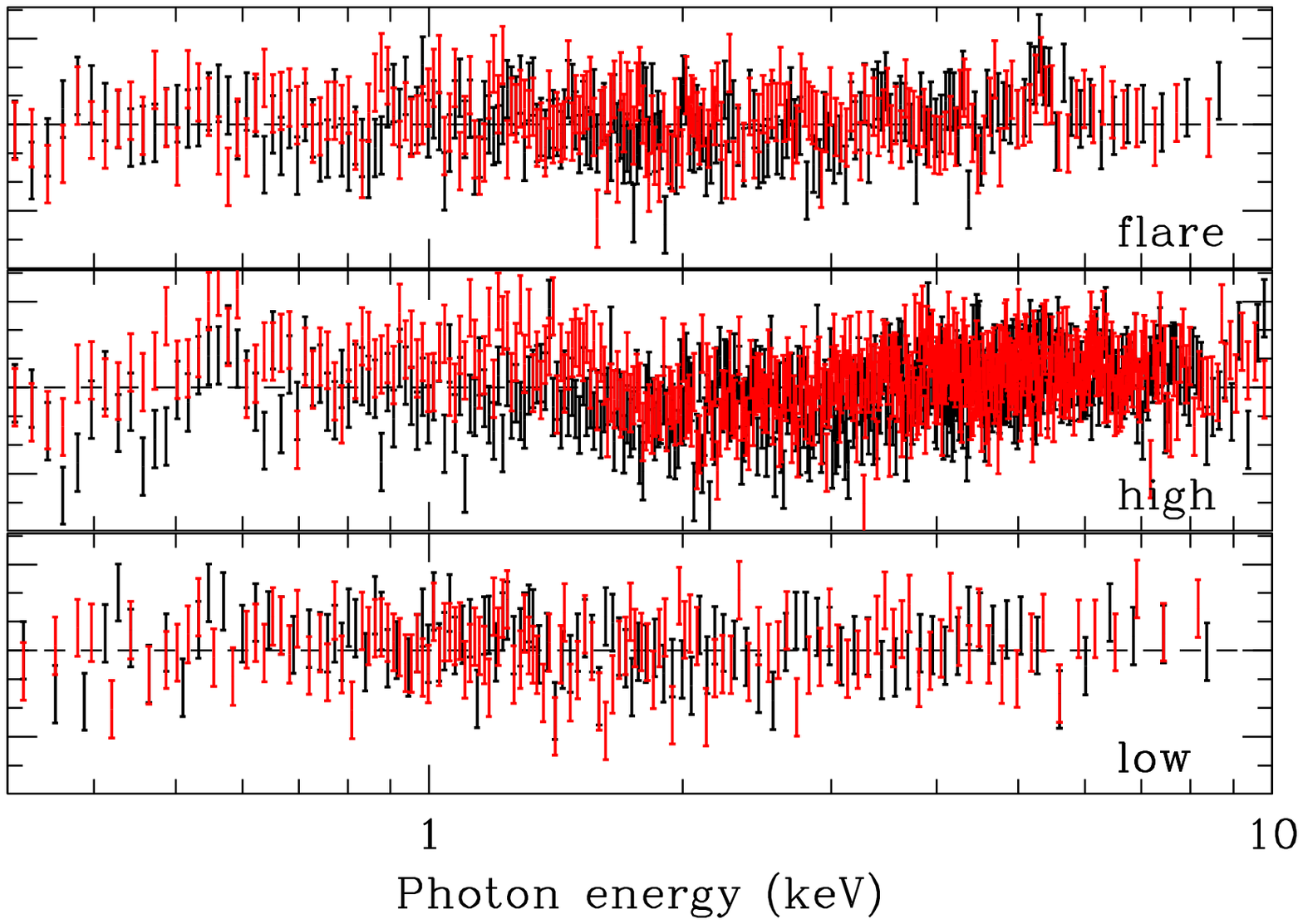}
\end{center}
\caption{\textit{XMM-Newton} MOS1 (black) and MOS2 (red) spectra from
  2013 November (left) and 2014 June (right) of PSR J1023+0038 in the
  three distinct flux modes: flare, high, and low. In all
  instances, an absorbed power-law model is fitted. The bottom
  three panels show the best fit residuals for each mode. Note the
  broad residuals of the high mode spectra. See Table 3 for the best
  fit parameters.}
\label{fig:xspectra}
\end{figure*}

\section{X-ray Pulsations}
As reported in \citet{Arch14}, the \textit{XMM-Newton} data revealed
coherent modulations at the pulsar spin period in both the 2013
November and 2014 June observations.  Intriguingly, the pulsations
occur just in the high flux mode, with a pulsed fraction of 8\%
(0.3--10 keV). The profile is clearly double-peaked, with a
$\sim$$180^{\circ}$ separation in rotational phase, with each peak
presumably corresponding to emission from one of the diametrically
opposite magnetic polar caps of the neutron star.

Here we examine the energy dependence of the high mode X-ray pulse
profiles as well as any spectral variations as a function of spin
phase. The method for extracting the pulse profiles is given in
\citet{Arch14}.  The top right panel of Figure \ref{fig:pulse_prof}
shows the high mode pulse profile in four energy bands (0.3--0.7,
0.7--1.5, 1.5--3, and 3--10 keV) obtained by aligning the total
profiles from the two individual observations via cross-correlation
\citep[see][for details]{Arch14}. The hardness ratio of the high mode
exhibits a subtle spectral softening in the trailing edge of the
stronger pulse.  For reference, we also show the flare and low mode
data folded using the same ephemeris, where no statistically
significant pulsations are seen.  The strongest, albeit marginal (with
a single-trial significance of only $\sim$2$\sigma$) signal is found
above 3 keV for the flare mode, although it shows only single-peaked
modulation.

As shown in Figure~\ref{fig:pulse_energy}, the spectrum of the
pulsations is similar to the spectrum of the source generally. This
suggests that the pulsed and most of the unpulsed radiation are
produced by the same process.  It is possible that the pulsations
decline above $5$\,keV but this cannot be conclusively established due
to the limited photon statistics above this energy.

\section{X-ray Spectroscopy}
Previous studies of PSR J1023+0038 have established that the X-ray
emission in the LMXB state is well-described by an absorbed power-law
with spectral photon index of $\Gamma\approx1.7$ with no substantial
spectral changes despite the over one order of magnitude variation in
flux \citep{Pat14,Tak14,Cot14,Ten14}.  The large photon harvest of the
\textit{XMM-Newton} data presented here allows us to identify any
subtle differences in spectral properties between the three distinct
flux modes: low, high, and flare. The corresponding spectra were
obtained using selections based on the count rate ranges for the low
and high modes determined in \S4, while for the flares we choose a
count rate cut of $\ge$$15$ counts s$^{-1}$ to minimize contamination
from the high mode.  Due to the relatively large uncertainties in the
spectral calibration of pn timing mode data \citep[see,
  e.g.,][]{Arch10,Bog13}, we only consider the MOS1/2 spectra in this
analysis.

%
%
\begin{figure*}[!t]
\begin{center}
\includegraphics[width=0.92\textwidth]{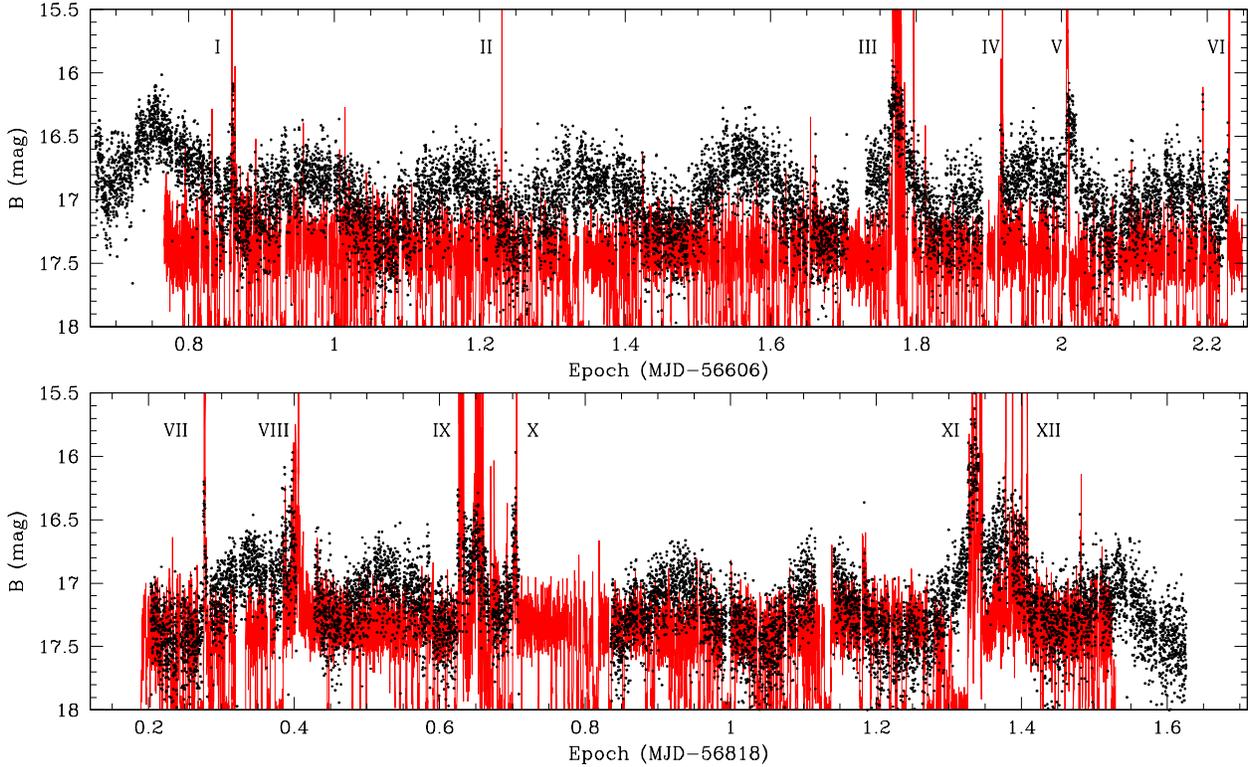}
\end{center}
\caption{\textit{XMM-Newton} OM fast mode light curve of PSR J1023+0038
  in the $B$ filter from 2013 November (top) and 2014 June
  (bottom). Each black point corresponds to a 10-s exposure. The
  error bars are omitted for the sake of clarity but are typically
  $\pm$0.2 mag. For reference, the 0.3--10 keV X-ray light curve (red)
  is plotted to show the relative alignment between the X-ray and
  optical flares. Gaps in the optical data are due to interruptions in
  exposure.}
\label{fig:om}
\end{figure*}

%
%
\begin{figure}[!t]
\begin{center}
\includegraphics[width=0.455\textwidth]{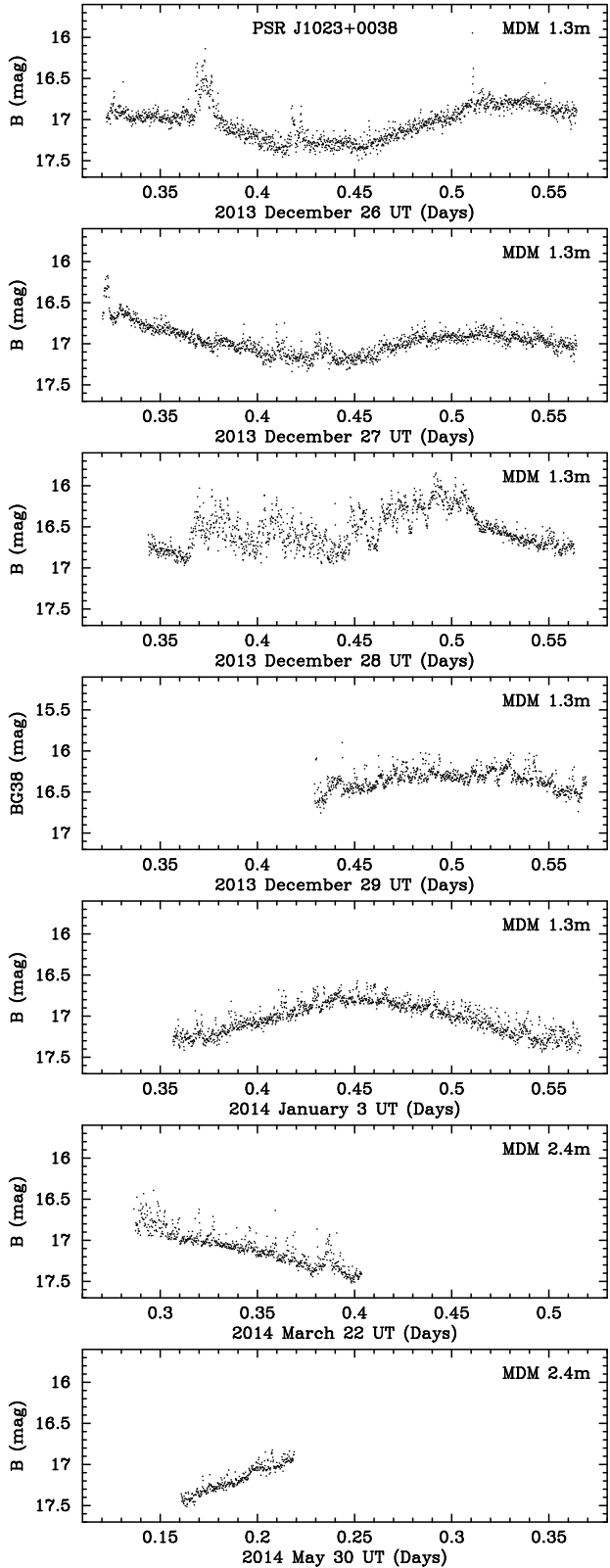}
\end{center}
\caption{MDM light curves of PSR J1023+0038 with 13~s cadence, plotted
  as a function of heliocentric UT date.  The orbital period of the
  binary is 0.198~d.  Magnitudes are based on differential photometry
  with respect to a calibrated comparison star, as described in \S2.2.3.}
\label{fig:mdm}
\end{figure}

Table 4 summarizes the results of the power-law spectral fits for the
three flux modes fitted in {\tt XSPEC} \citep{Arnaud96}. We conducted
the analysis in two ways: one with each mode fitted separately and the
other with all three modes fitted jointly with a tied value of the
hydrogen column density along the line of sight ($N_{\rm H}$). In both
instances, the fits suggest that the value of $N_{\rm H}$ is
consistent among the three with no appreciable enhancement in
the intervening absorbing column within the binary during any of the
modes.

%
%
\begin{figure*}[!t]
\begin{center}
\includegraphics[width=0.85\textwidth]{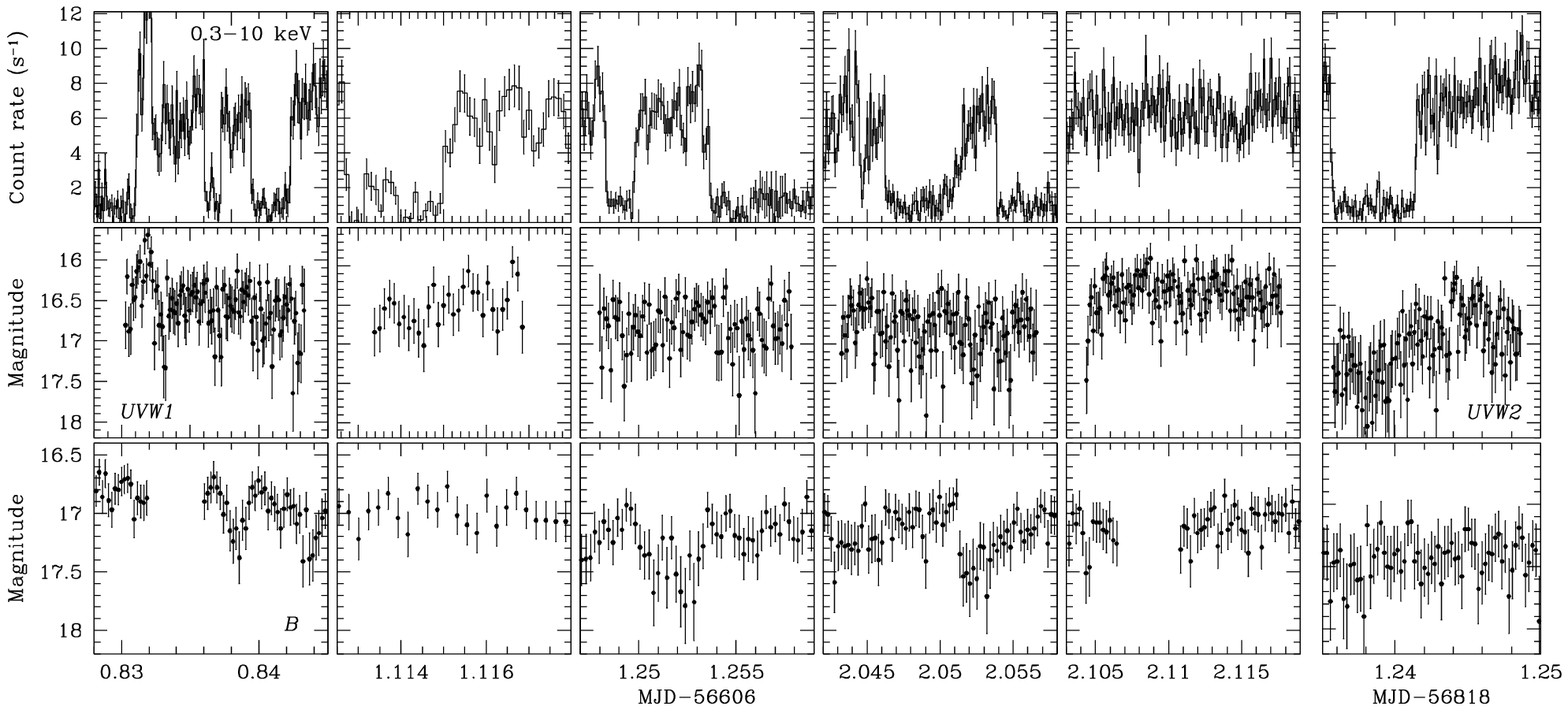}
\end{center}
\caption{Concurrent \textit{XMM-Newton} EPIC X-ray (top),
  \textit{Swift} UVOT UV (middle), and \textit{XMM-Newton} OM optical
  (bottom) light curves. All but the right-most set of panels show
  observations during 2013 November 10-12 when the \textit{Swift} UVW1
  was used. The right-most panels show the data during 2014 June 11,
  when the \textit{Swift} UVOT UVW2 filter was in place.}
\label{fig:uvot}
\end{figure*}

%
%
\begin{figure}[!t]
\begin{center}
\includegraphics[width=0.47\textwidth]{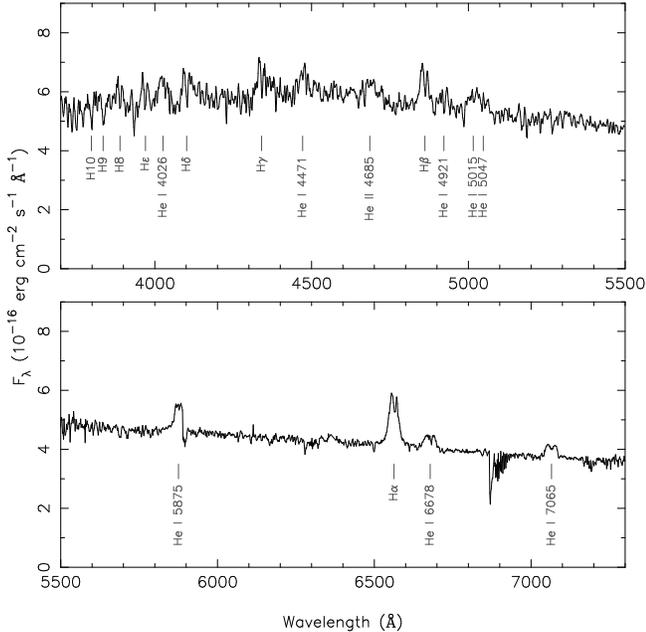}
\end{center}
\caption{VLT X-Shooter spectrum of PSR J1023+0038 from 2013 November 12.}
\label{fig:vlt}
\end{figure}

For the flare and low modes, the emission is well-described by an
absorbed power-law in both observations. In contrast, the fits to the
high mode spectra yield null hypothesis probabilities of
$3\times10^{-6}$ and $2\times10^{-9}$ for the 2013 November and 2014
June data, respectively. This means that a simple absorbed power-law model
alone does not provide an adequate representation of the high mode
spectrum, which is also evident from the broad-band residuals shown in
Figure \ref{fig:xspectra}. One likely explanation is the presence of a
faint thermal (e.g., neutron star atmosphere) component since the
X-ray pulsations imply accretion onto the neutron star, which may
result in at least some superficial heating of the surface.  To
account for this emission we consider two models: {\tt nsa}, a
passive, weakly magnetized neutron star hydrogen atmosphere
\citep[][]{Zavlin96} and, {\tt zamp}, a hydrogen atmosphere accreting
at a low rate \citep[][]{Zamp95}. For the latter, the spectrum is
defined in terms of the observed accretion luminosity expressed in
terms of the Eddington rate, assuming radiation from the entire
surface of a neutron star with $M_{NS}=1.4$ M$_{\odot}$ and
$R_{NS}=12.4$ km.  The best fit parameters of the thermal component
are consistent between the 2013 November and 2014 June observations
(Table 5).  The implied effective emission radii of the {\tt nsa}
model are smaller than the whole neutron star, consistent with
emission from hot spots.  The implied contribution of thermal
radiation to the total luminosity ($3-9$\%) is consistent with the
pulsed fraction of the X-ray pulsations in the high mode. The implied
luminosity from the {\tt zamp} model is $2.5\times10^{-5}L_{\rm edd}$,
which translates to a thermal luminosity of $\sim$$6\times10^{33}$ erg
s$^{-1}$, much greater than the implied thermal fraction. This arises
due to the fact that the model considers emission from the entire
surface, whereas the pulsations from PSR J1023+0038 indicate emission
from a small portion of the surface.  Although not strictly correct,
adjusting the size of the emitting area to that obtained with the {\tt
  nsa} model produces a value consistent with the thermal luminosity
deduced from the spectral fits. While both composite models result
in a reduction of the broad residuals relative to the pure power-law
model, for both sets of observations the null hypothesis probabilities
($7\times10^{-3}$ and $1\times10^{-3}$, respectively) indicate
fits that are formally not acceptable.  It may be that the
flare and low state have similar residuals but due to the lack of
photon statistics they are not as apparent.

A possible explanation for the substantial residuals in the
high mode fits could be the presence of a complicated spectrum
produced by three or more distinct emission
components. However, adding components commonly used in
  modeling of LMXB spectra such as thermal Comptonization or
  bremsstrahlung does not produce acceptable fits either.

An alternative interpretation involves small fluctuations of the
power-law photon index during the high mode, which would result in a
mean spectrum that is not well represented by a power-law with a
single photon index. To investigate this possibility we have
  examined the ratio of counts in the 1.5--10 keV and 0.3--1.5 keV
  bands versus count rate using the combined pn and MOS1/2 data. We
  consider three count rate ranges within the high mode with bounds
  $4.7$, $6.2$, $7.8$, and $9.3$ counts s$^{-1}$, chosen to minimize
  the contamination from the low and flare modes (see
  Figure~\ref{fig:xmm_poisson_smoothed_histogram}). For the 2013
  November data, this hardness ratios are $0.666\pm0.003$,
  $0.678\pm0.003$, and $0.804\pm0.005$ at the lower end, peak, and
  upper end of the count rate distribution of the high mode,
  respectively. For the 2014 June data, the values are
  $0.637\pm0.002$, $0.606\pm0.002$ and $0.685\pm 0.005$, respectively.
  In both instances, the difference in hardness as a function of count
  rate is highly significant, which is an indication of spectral
  variability within the high mode. Moreover, for a given count rate
  the hardness ratio is not the same between the two observations with
  the spectrum being generally softer in the 2014 June data (as
  evident from the best-fit power-law indices in Table 4 and confirmed
  by the values of the hardness ratio). Therefore, fluctuations in the
  power-law index can explain the poor quality of the spectral fits.

The flux in the flare, high, and low modes appears to decrease by
11\%, 3.5\% and 16\% between 2013 November and 2014 June.  We have
investigated whether this could be due to the choice of count rate
cuts for the three flux modes. Using more conservative cuts produces
similar results for the flux and photon index.  Thus, the long-term
flux changes of the modes appear to be intrinsic to the system.

We note that for the case of the flare spectra the quoted
luminositites represent time-averaged values. The peak luminosity,
observed in flare IX, reaches a luminosity of
$\approx$$3\times10^{34}$ erg s$^{-1}$ (0.3--10 keV), assuming the
same spectral shape as the total flare spectrum.

\section{Optical/UV Variability}
As noted by Halpern et al.~(2013), in the optical the PSR J1023+0038
binary is $\sim$1 magnitude brighter during the LMXB state than in the
radio pulsar state.  From the \textit{XMM-Newton} OM and MDM
photometric light curves (Figures~\ref{fig:om} and \ref{fig:mdm})
orbital modulation similar to the heating light curve observed during
the radio pulsar state \citep{Thor05,Woudt04} are also
evident. However, the variations now appear more sinusoidal and
symmetric about the peak.  The optical brightness varies between $B
\approx 17.5$ and $B \approx 16.7$, although orbit-to-orbit evolution
of the overall brightness is apparent. For comparison, in the radio
pulsar state the $B$ filter magnitude varies between $18.4$ and $17.8$
\citep{Thor05,Homer06}.

A wide variety of flaring behavior is also seen, which is associated
with the accretion disk state.  Some of these flares are very rapid,
lasting less than 1 minute, with a rise time that is unresolved at our
10~s (OM) and 13~s (MDM) cadences.  A good example of this phenomenon
can be seen in the MDM light curve on December 26.51 UT.  At the other
extreme is a continuous episode of strong flaring on December 28 that
lasts 3.6~hr, or 3/4 of the binary orbit.  Low-level flaring is
present almost continuously for the entire 2013 November
\textit{XMM-Newton} observation and the MDM 2013 January 3
observation.  We do not see any optical moding behavior that resembles
the X-ray variability.  In all cases, the optical variability can be
characterized as positive flares superposed on the otherwise smooth
heating light curve.

The $B$ filter photometric data obtained with the \textit{XMM-Newton} OM
further reveal that the brightest optical flares closely match the
prominent X-ray flares.  Therefore, the flares in PSR J1023+0038
appear to be broad-band phenomena that span from the optical up to at
least hard X-rays \citep[as seen with \textit{NuSTAR};][]{Ten14}.  Of
particular note are flares III and XI, which reach peak $B$ filter
magnitudes of 15.9 and 15.5, respectively. For some, especially flares
I and V, the flux peak occurs at significantly later times than the
X-ray one and the optical tail of the flare is much longer (see
Figure~\ref{fig:flares}). For the brief but intense X-ray flare
labeled as II, there does not appear to be a corresponding optical
event, while multiple fainter rapid X-ray flares (with X-ray count
rates below 20 s$^{-1}$) seen in Figures 2 and 3 have clear optical
counterparts.

To identify any time lags between the X-ray and optical flares, we
cross-correlated the \textit{XMM-Newton} X-ray and optical time series
binned at 10 s resolution. We first fitted a sine function to the OM
observation and subtracted it from the data to remove the orbital
modulation. The best-fit sine function results in a period of 4.754 h,
consistent with the orbital period of the system. As expected, the
cross-correlation between the X-ray and detrended optical data sets
reveals a strong correlation, which is dominated by the flares. The
time lags among the flares range between -260 s and +160 s, where a
positive lag indicates that the X-ray flare precedes the optical one.
The strongest correlation is found for the short flares I (with time
lag +130 s) in 2013 November and X (with lag 0 s) in 2014 June. For
flares III, VIII, IX, and XI the optical leads the X-ray (by -160,
-260, -100, -30 s, respectively) although the correlation is weak.
The time lags of individual flares are unlikely to be correlated with
the orbit due to the variation of the emission site across the orbit
with respect to the observer because the compactness of the binary
($\sim$4-5 light-s). Indeed, we find no clear trend of the time-lag
changes with the orbital phase.  Finally, we investigated any
correlation of the low level flux variation between the X-ray and
optical by removing all significant flares.  The resulting
correlations are very weak, consistent with the apparent lack of flux
moding in the optical photometric data.

Another way to reveal low-level optical flux variations
correlated with the X-ray mode switching is to consider the optical
emission averaged over many low and high X-ray mode intervals.  To
this end, we first removed the sinusoidal variations of the optical
time series.  For each X-ray low mode we then computed averaged $B$
filter light curves for specific durations (100, 200, and 300 s) and
compared them against the high mode immediately before and after using
the same duration.  The 2013 November data show that the source is on
average $\sim$0.1 mag brighter during the X-ray low modes relative to
the high modes that occur immediately before or after. However, the
2014 June data does not show the same behavior.

The UV emission observed with the \textit{Swift} UVOT (see
Figure~\ref{fig:uvot}) exhibits rapid variability as well.
A moderately bright flare that occurs around MJD $56606.83$ has an
obvious UV counterpart in the \textit{Swift} UVOT data (as seen in the
left-most panels of Figure~\ref{fig:uvot}).  The UVOT $UVW1$ filter
data from 2013 November and the $UVW2$ filter data from 2014 June
cover time periods when multiple low flux intervals are seen in the
\textit{XMM-Newton} data. There are no obvious $UVW1$ brightness
variations commensurate with the nearly order of magnitude X-ray flux
changes associated with the mode transitions.  The 2014 June $UVW2$
data exhibits a brightness increase that coincides with a low-high
X-ray mode transition, although due to the brief UVOT exposure it is
not clear if this flux variation occurs consistently for mode
transitions.

\section{Optical Spectroscopy}

In Figure~\ref{fig:vlt}, the average of the two individual VLT spectra
is plotted. The spectrum is dominated by a blue continuum and strong,
double-peaked emission lines of H and He. The spectrum appears similar
to other optical spectra of the PSR\,J1023+0038 binary presented in
the literature, both during the active period in 2000/2001
\citep{Bond02,Wang09} and 2013/2014 \citep{Halpern13,Tak14,Cot14}. The
optical spectra confirm that the accretion disk was present shortly
after the end of the 2013 November X-ray observation. As the optical
spectra presented in the literature since the 2013 June transition all
show the presence of an accretion disk, it is highly probable that it
has been present throughout the 2013 November and 2014 June X-ray
observations of PSR\,J1023+0038. A detailed analysis of these and a
series of other spectra acquired since the state transformation of PSR
J1023+0038 will be presented in a subsequent paper.

%
%
\begin{figure*}[!t]
\begin{center}
\includegraphics[width=0.45\textwidth]{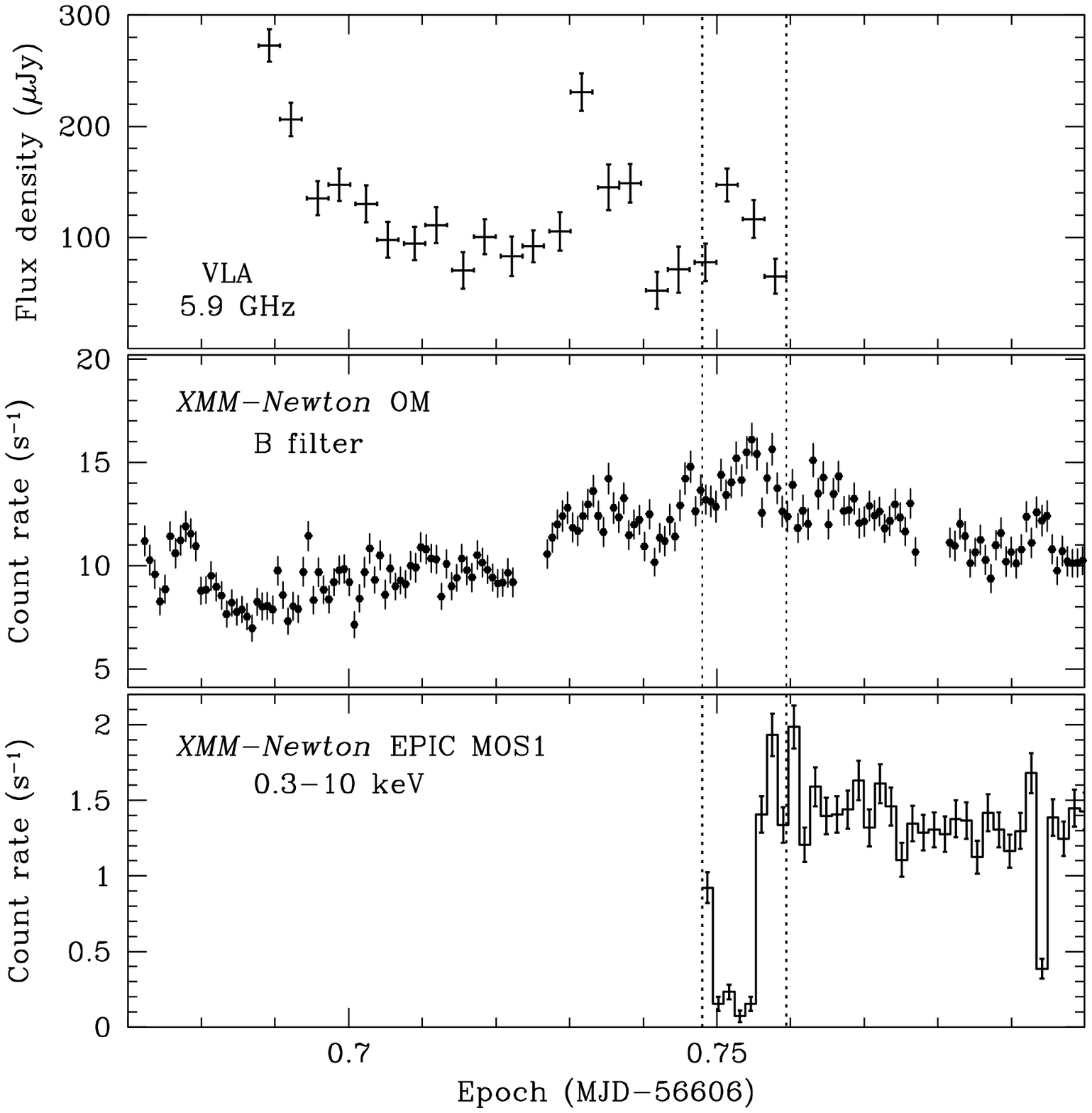}
\includegraphics[width=0.256\textwidth]{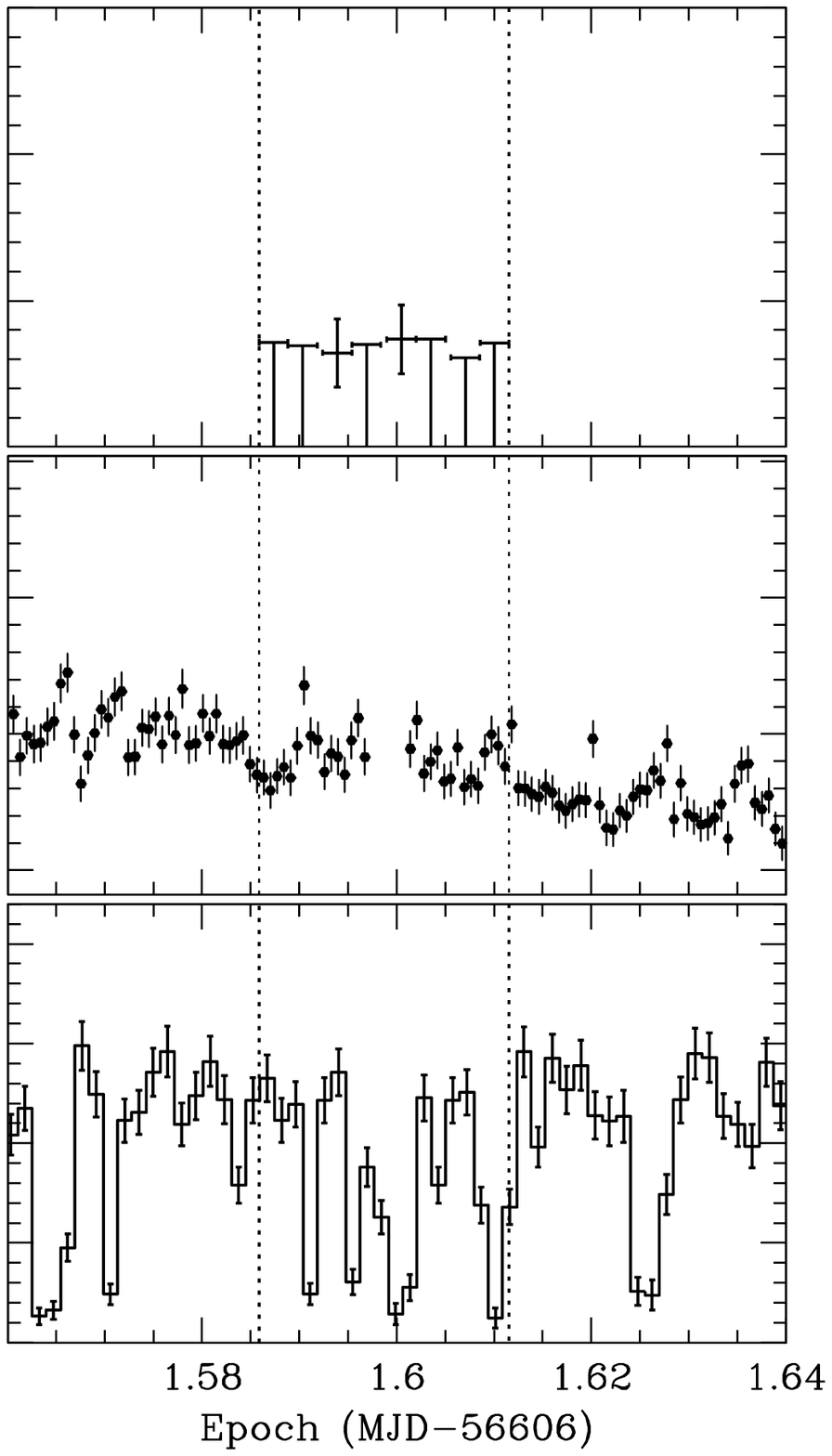}
\end{center}
\caption{Top: VLA 5.9 GHz light curves of PSR J1023+0038 from 2013
  November 10 and 11. Middle: \textit{XMM-Newton} OM light curve.
  Bottom: \textit{XMM-Newton} EPIC MOS1 light curve of PSR
  J1023+0038. The vertical dotted lines demarcate the time overlap of
  the radio and X-ray observations.}
\label{fig:jvla}
\end{figure*}

\section{Radio Pulsation Search} 
As reported in \citet{Stap14}, the 2013 June reappearance of the
accretion disk in PSR J1023+0038 was accompanied by cessation of its
bright radio pulsations. To verify this using the 2013 November and
2014 June observations, the radio data were all folded with the
ephemeris derived prior to the pulsar disappearance in 2013 June. In
both cases sub-integration times of 10\,s were used with 1600
frequency channels of with 0.25 MHz for the LT and 512 frequency
channels of width 0.3125 MHz for the WSRT data. No radio pulsations
were detected around the pulsar period.

Conceivably, if the rotation-powered pulsar emission mechanism is
actually still active, the radio pulsations may be highly intermittent
such that they may only be present in one of the three distinct X-ray
flux modes. In such a scenario, the radio emission is either quenched
or severely obscured in the other modes and is only able to ``break
through'' during one mode.  To investigate this possibility, we
divided the radio observations into time segments that match the times
of the high, low, and flare modes observed during the simultaneous
\textit{XMM-Newton} observations. For this purpose, we defined an
X-ray low mode when the count rate is less than 3\,counts s$^{-1}$ and
the duration of the low mode is more than 40\,s. This resulted in 70
and 35 low mode intervals overlapping with the LT and the WSRT data,
respectively. The data in these sections were then corrected for the
new, local ephemerides derived from the X-ray data
\citep[from][]{Arch14} and subsequently a search in period and
dispersion measure (DM) was undertaken for each section. The period
search range was 0.01 $\mu$s and the DM search range was
3\,cm$^{-3}$\,pc. These values were chosen to allow for significant
changes in the orbital parameters and/or in the electron column along
the line of sight due to material being lost from the companion, or
due to the accretion disk. To a signal-to-noise limit of 6 we find no
evidence of pulsed emission during any of the low mode intervals in
either the LT or WSRT data. We note that these data span a wide range
of orbital phases when the pulsar has previously been detected with
signal-to-noise ratios of many tens in similar integration times. We
performed the same analysis on flares that occur during the radio
observation with no significant detection of pulsations
either. Finally we considered a long sequence of high mode X-ray
emission (MJD 56607.348--56607.385) overlapping with the LT data and
also find no significant pulsed signal.

\section{Radio Imaging}
The two-hour long VLA observation made at 5 and 7 GHz at the beginning
of the 2013 November campaign shows a strong detection, with a flat
spectrum ($\alpha = 0.09 \pm 0.18$, where flux density is proportional
to $\nu^\alpha$) and flux density which varies rapidly between 60 and
280 $\mu$Jy (see left panel of Figure~\ref{fig:jvla}). In the 5 GHz e-EVN VLBI
observations made shortly after the 2013 November observations we
obtain a marginal detection \citep[3.8$\sigma$ peak with flux density
  50 $\mu$Jy at the precise position of the pulsar predicted by the
  astrometric solution of][]{Del12}; the much sparser $uv$ coverage
and the faintness of the source precluded any detection of variability
on a sub-observation timescale.  In the shorter 2--4 GHz VLA
observation made 20 hours later, the source is not detected in the
combined dataset, with a 3$\sigma$ upper limit of 30\,$\mu$Jy.  When a
light curve is made with a resolution of 4 minutes, two tentative,
3$\sigma$ detections with peak flux density $\sim$60\,$\mu$Jy are made
(right panel of Figure~\ref{fig:jvla}).  Finally, PSR J1023+0038 is
not detected in the 140 MHz LOFAR observations, although the upper
limit of several mJy is not constraining (we note however that, when
visible, the pulsar has a period-averaged flux density of $\sim$$40$
mJy at 150MHz; Kondratiev et al., in prep).  \citet{Del14} show
that the variable, flat-spectrum emission seen here persists over many
observations made in a six-month period after the November 2013
observations.  The observed radio emission is inconsistent with a
radio pulsar origin, which should be very steep \citep[$\alpha \sim
  -2.8$;][]{Arch09} and hence much brighter than the observed emission
below 4 GHz, whilst being visible up to $\sim$6 GHz.

Only the last 16 minutes of the VLA data from 2013 November 10 overlap
with the \textit{XMM-Newton} EPIC MOS1 exposure\footnote{Due to
  different instrumental overheads, the three EPIC X- ray detectors do
  not have the same exposure start times.}. During most of this period
the X-ray emission is in the low flux mode, while the radio emission
exhibits a rise and fall in flux; the overlap is too brief to
establish whether any correlation (potentially with a time lag due to
different production sites for the radio and X-ray emission) exists.
Although the \textit{XMM-Newton} OM data cover most of the VLA
observations, no discernable relation between the radio and optical
variability is present.  For the 2013 November 11 radio observation,
the faintness of the radio emission makes it impossible to identify
any relation between the radio and X-ray emission.

Although it is thus difficult to establish whether there is any
correlation between the radio and X-ray light curves, we can still
examine whether there are any noticeable differences in the radio
emission during the different X-ray modes.  As discussed in \S11.1,
this has significant implications for whether the radio pulsar could
still be active at any time during the current LMXB state, since the
activation of the radio pulsar would lead to the presence of
additional steep-spectrum radio emission.  A significant portion of
the 16 minute overlap between the MOS1 and VLA observations on
November 10 is in the low mode, and when we image the VLA data from
18:00:00UT to 18:06:30UT (during this mode) separately, we obtain a
flux density of $180 \pm 15 $ $\mu$Jy with a spectral index of $0.4
\pm 0.6$: consistent with the average value for the whole observation.
During the 2--4 GHz observation on 2013 November 11, multiple examples
of the low mode are visible, generally for a period of a few minutes
at a time. In a radio light curve with 1 minute resolution, no peaks
$>$2.5$\sigma$ ($\sim$100 $\mu$Jy) were visible.  In either of these
cases (the 6.5 minute low mode image in the November 10 observation
and the 1 minute time slices in the November 11 observation) the radio
pulsar emission, if unchanged compared to that reported in
\citet{Arch09}, would have been clearly visible.  We discuss the
implications in \S11.1.

\section{Discussion}

Our unprecedented multi-wavelength campaign of PSR J1023+0038,
conducted in 2013 November and 2014 June, shows fascinating
phenomenology that provides fresh insight into the properties of PSR
J1023+0038 and analogous systems.  Based on the wealth of data we have
accumulated, we are able to establish the following.

In its LMXB state, PSR J1023+0038 resides in a luminosity range of
$L_X\sim 10^{32-34}$ erg s$^{-1}$ (0.3--10 keV). The majority of the
time is spent at a $3\times 10^{33}$ erg s$^{-1}$ level, during which
coherent X-ray pulsations are observed. This implies that active
accretion onto the stellar surface takes place \citep{Arch14}.  During
the low flux mode intervals ($\sim$$5\times 10^{32}$ erg s$^{-1}$), no
pulsed emission is seen, possibly because the accretion flow is unable
to reach the star.  As shown in \citet{Arch14}, there is no evidence
for pulsations in the low flux mode, with a 95\% confidence upper
limit on the pulsed fraction of $\sim$2.4\%. This excludes the
possibility that the low mode simply has the same amplitude of
pulsations but with scaled-down luminosity. Instead, it appears that
the coherent X-ray pulsations are either completely absent or have a
comparatively very low amplitude. 

The low mode 0.3--10 keV luminosity is $\sim$5 times greater than the
average luminosity observed in the radio pulsar state in the same
energy band ($9\times10^{31}$ erg s$^{-1}$). In the radio pulsar
state, PSR J1023+0038 showed evidence for broad double-peaked
pulsations at the spin period at a 4.5$\sigma$ with a $\sim$9\% pulsed
fraction for 0.3--10 kev \citep{Arch10}, in addition to the
dominant shock emission component. Such pulsations are observed in
multiple rotation-powered MSPs and are believed to be produced by
heating of the magnetic polar caps by a return flow of relativistic
particles from the pulsar magnetosphere.  If the X-ray pulsations seen
in the radio state are present in the low mode they
would appear with a pulsed fraction of $\sim$2\%, below the upper
limit we have derived. Therefore, we cannot rule that the same X-ray
pulsations seen in the radio state are present in the low mode of the
LMXB state.

\citet{Arch14} demonstrated that the pulsations appear to be
completely absent during the flares as well. The pulsed fraction upper
limit of 1.5\% corresponds to a pulsed flux of 0.21 counts s$^{-1}$,
lower than the pulsed flux in the high flux mode (0.33 counts
s$^{-1}$).  This indicates that during the flares the pulsations are
strongly supressed. Therefore, the flares are not simply added flux on
top of the high mode emission as the pulsations would still be easily
detectable. Instead, the high mode appears to be absent during the
flares.

The multiple intense X-ray flares exhibit varied morphologies and
temporal behavior.  Flaring/burst activity in accreting systems might
be from coronal flares arising due to magnetic reconnection events in
the accretion disk \citep[e.g.,][]{Gal79}.  Alternatively, an
instability in the inner disk may cause a large influx of acreting
material towards the compact object (see \S11.4). Of particular
interest are the long flares interspersed with rapid drops in flux
(e.g. III an IX in Figure~\ref{fig:flares}) down to X-ray flux levels
comparable to the low mode.  This suggests that at least for these
flares the emission is not superposed on the high mode flux, which
offers an additional line of evidence for the lack of high mode flux
in the flares.

There is no indication for the orbital-phase-dependent X-ray
modulations that were seen in the radio pulsar state even if all
intervals of a single flux mode are folded at the binary period. The
non-thermal spectrum in each of the three flux modes is much softer
compared to that seen in the radio pulsar state, which had
$\Gamma=1-1.2$ \citep{Arch10,Bog11}.  This implies that the
intra-binary shock near the face of the companion, responsible for the
X-ray emission in the disk-free state, is either absent or completely
overwhelmed in the accreting state.  The change in the optical heating
light curve from an asymmetric to a symmetric, more sinusoidal shape
favors the absence of this shock emission in the LMXB state. This
difference can be explained if the heating by the pulsar wind was
asymmetric due to channeling of the pulsar wind by the companion's
magnetic field \citep[see, e.g.,][]{Li14b}, while the heating in the
LMXB state is by X-rays from the inner disk, which would produce a
more symmetric light curve as the source of heating is more
  point-like.  In addition, even if the rotation-powered pulsar
mechanism is still active in the LMXB state, the accretion flow may
intercept a substantial fraction of the pulsar wind directed towards
the companion \citep[see, e.g.,][]{Tak14}.

The sudden drops to a low flux mode appear to be of an entirely
different nature than the dips in the X-ray dipper variety of LMXBs
\citep[e.g.,][]{White82,Smale88,Balu12} as they do not show spectral
changes in either soft X-rays (as shown in \S6) or hard X-rays
\citep[see][]{Ten14}.  In general, based on the results of the X-ray
spectral fits presented in \S6, any interpretation that evokes
obscuration/absorption is unlikely. Since there are negligible
spectral changes, especially the value of $N_{\rm H}$ (see Tables 3
and 4), between the high and low modes, there is no indication for an
enhancement in the amount of intervening material during the low mode.
There is also no evidence of an X-ray luminosity dependence on the
duration and frequency of flares and low flux mode intervals or any
correlation between the separation between (and duration of) dips or
flares.

%
%
\begin{figure*}[!t]
\begin{center}
\includegraphics[width=0.75\textwidth]{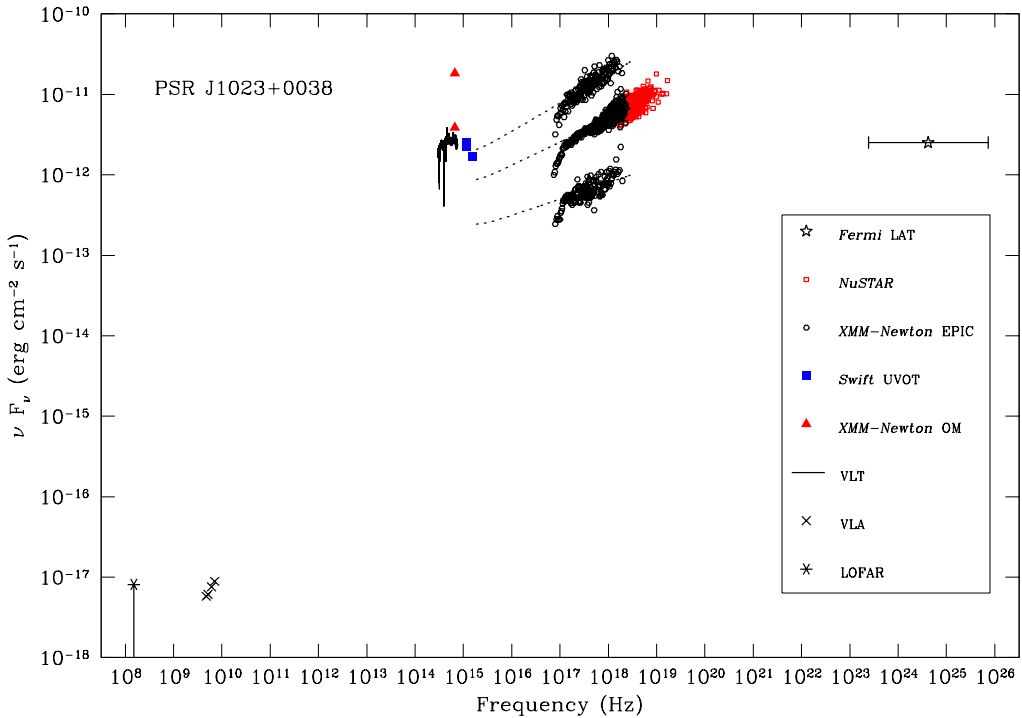}
\end{center}
\caption{The broad-band spectrum of PSR J1023+0038 in the LMXB state
  spanning from the radio to the $\gamma$-ray range. The red triangles
  show the median and maximum observed fluxes in the $B$ filter as
  observed with \textit{XMM-Newton} OM, while the blue squares show
  the mean UV fluxes observed with \textit{Swift} UVOT. The dotted
  lines show the extrapolation of the best-fit power-law spectra
  fitted to the flare, high, and low X-ray flux modes observed with
  \textit{XMM-Newton} EPIC (open circles). The \textit{Fermi} LAT flux
  in the 1--300 GeV range is based on the value reported in
  \citet{Stap14}, while the \textit{NuSTAR} spectrum is taken from
  \citet{Ten14}.}
\label{fig:broad}
\end{figure*}

Although the extensive data set presented here reveals a wide variety
of interesting flaring phenomenology there is evidence that it does
not cover the whole range of possible behavior of PSR J1023+0038 in
the LMXB state. In particular, in the 100 ks \textit{NuSTAR}
observation of PSR J1023+0038 from 2013 October, the system undergoes
a prolonged episode of flaring activity that lasts for $\sim$10 hours
\citep{Ten14}. For reference, the longest flare in the
\textit{XMM-Newton} observations last for $\sim$45 minutes and the
longest optical flare seen in the MDM data (from 2013 December 28) is
$\sim$3.6 hours long.  It is unclear if this long flare is an
exceptionally long and bright flare episode or that it is yet another
flux mode.

As shown in \citet{Cot14} and confirmed by our analysis, the optical
emission exhibits orbital modulations, which arises due to heating of
the face of the secondary star. Sporadic intense flares correlated
with the X-ray ones are also seen in both the UV and optical.  Neither
the UVW1 or B filter photometric light curves exhibit large-amplitude,
rapid flux mode switching seen in X-rays.

Based on the multi-wavelength data we have presented here, we can
investigate the emission properties of PSR J1023+0038 in its present
accreting state spanning the electromagnetic spectrum from the radio
to the GeV $\gamma$-ray range. Figure~\ref{fig:broad} shows the
spectrum of PSR J1023+0038 across nearly 18 decades of photon
frequency.  The X-ray spectrum is separated into the three flux
modes. In \citet{Ten14} it was established that the hard X-ray
emission observed with \textit{NuSTAR} exhibits the same power-law
spectrum and mode switching behavior as the soft X-rays. For
reference, here we show the time-averaged hard X-ray spectrum that is
also dominated by the high mode, as well as the \textit{Fermi} LAT
flux measurement for PSR J1023+0038 in the LMXB state from
\citet{Stap14}.

One important aspect of the observed emission is how the X-ray
spectrum in the low, high, and flare modes extrapolates to the UV and
optical range. From Figure~\ref{fig:broad} it is evident that the
emission process responsible for the X-ray flux cannot account for the
majority of the optical emission as observed with \textit{XMM-Newton}
OM (even if we remove the sinusoidal modulations).  This provides
a plausible explanation for the absence of large-amplitude variations
in the $B$ filter data comparable in amplitude to the X-ray mode
changes. The extrapolated high mode spectrum in the $B$ filter
contributes $\lesssim$10\% to the average optical flux.
The high mode spectrum can produce $\sim$50\% of the observed emission
in the \textit{Swift} UVOT UVW2 filter, while in the UVW1 filter the
contribution of the high mode declines to $\sim$30\%. The increase in
UVW2 emission that appears to coincide with a low-high mode transition
and the lack of comparable brightness variations in the UVW1 band
during X-ray mode switches are consistent with these findings.

The extrapolated flare mode spectrum falls well below the flux of the
optical flares, suggesting that the X-ray and optical flares are
possibly not generated by the same process. Energetically, the peak
optical luminosity in the brightest flare ($\sim$$1\times10^{34}$\,erg
s$^{-1}$) is well below the peak X-ray luminosity
($\sim$$3\times10^{34}$ erg s$^{-1}$ for 0.3--10 keV or
$\sim$$6\times10^{34}$ erg s$^{-1}$ for 0.3--79 keV if we take into
account the \textit{NuSTAR} data) so it is plausible that reprocessing
at a location other than the site of X-ray emission produces the
optical flares.  However, reprocessing cannot easily account for the
puzzling instances where the optical peak precedes the X-ray one. The
projected size of the binary is $\sim$4--5 light-seconds so these
significant time offsets cannot be attributed to light travel time
differences due to emission from different regions of the system. A
possible explanation may involve a complex emission spectrum that peaks in
both the optical and X-rays.

\subsection{Radio Pulsar Enshrouding or Quenching?}
It has been suggested that the radio pulsar may be active in the LMXB
state \citep[e.g.,][]{Tak14}. While we believe the detection of X-ray
pulsations is evidence that the radio pulsar is quenched in the high
mode (but see \S 11.2), it remains possible that the low and/or flare
modes corresponds to times during which the magnetosphere is free of
material and the radio pulsar becomes active. Our non-detection of
radio pulsations in this mode shows that either the radio pulsar is
quenched in the low mode as well, or it is hidden by temporal
scattering or absorption.  The latter two possibilities can be
addressed by our radio imaging observations.

As shown in \S10, the continuum radio emission from PSR J1023+0038 is
highly variable on timescales of minutes.  It also exhibits a spectrum
substantially flatter than the radio pulsar spectrum that extends at
least up to 18 GHz \citep[][]{Del14}.  If the radio pulsar was still
active but enhanced scattering due to the extra intervening material
had obscured the pulsations, the pulsar would still be visible as a
steep-spectrum source (potentially with a low frequency turnover due
to free-free absorption, as discussed below). There is therefore some
other mechanism producing the variable, flat-spectrum continuum
emission we see.  Neutron stars accreting from a binary companion
sometimes exhibit flat-spectrum radio emission, which is assumed to
originate from partially self-absorbed synchrotron radiation generated
in a collimated, jet-like outflow \citep[e.g.,][]{Mig06,Mig11}.  The
radio continuum behavior we observe is consistent with the jet-powered
synchrotron emission seen in other accreting neutron star systems
\citep[see][for further details]{Del14}.

This conclusion does not immediately address the possibility of the
radio pulsar mechanism operating intermittently during the low or
flare modes.  However, as noted in \S10, the contemporaneous radio and
X-ray light curves (Figure \ref{fig:jvla}) reveal that there are no obvious
changes in either radio flux density or spectral index when comparing
the X-ray high mode and low mode time ranges during the 2013 November
10 VLA observation.  Likewise, no bright, steep spectrum source
appears during the low mode intervals in the 2013 November 11 VLA
observation.  Therefore, we can immediately rule out scattering alone
as a cause of the non-detection of the radio pulsar emission.

The potential impact of free-free absorption is more difficult to
exclude.  In the radio pulsar state, eclipses due to absorption are
seen at frequencies up to several GHz \citep{Arch09,Arch13}, showing
that it is possible even for the relatively tenuous intervening
material present in this state to absorb the pulsar signal at low
frequencies. In the radio pulsar state the eclipses disappear
completely at frequencies of 3 GHz and above, meaning our imaging
observations would have easily detected pulsar emission if the
absorbing conditions were unchanged in the LMXB state.  However, even
a modest increase in the density (factor of a few) or length scale
(factor of 10) of the absorbing material would increase the free-free
optical depth sufficiently to hide any pulsar emission from our radio
continuum observations as well as our radio-pulsar-mode observations
at frequencies up to $\sim$ 6 GHz.  The necessary changes could be
driven by, e.g., the increase in heating of the companion, or
additional material emanating from the accretion disk.

Exact limits are difficult to place because of the additional
dependence on the temperature of the absorbing material, as well as
the fact that the spectrum of the pulsar is difficult to measure (the
value of $-2.8$ from \citet{Arch09} is particularly steep, even for a
millisecond pulsar, and the potential contamination by scintillation
means it should be applied here with caution; a shallower radio pulsar
spectrum would mean that the radio pulsar emission would still be
visible to the VLA up to $\sim$10 GHz).  In light of these caveats,
we cannot categorically rule out an active radio pulsar in the low
mode, and we note that a similar scenario --- active radio pulsar
enshrouded by absorption --- has been proposed for SAX J1808.4--3658
in quiescence \citep{Camp04}.  However, we note that any activation of
the radio pulsar in the low mode, even if the radio emission is
free-free absorbed, would also have to occur without substantially
affecting the ongoing flat-spectrum (jet) emission.

\subsection{Pulsar Mode Switching?}
An alternative explanation for the observed X-ray pulsations is
possible: the pulsar emission mechanism in some cases produces
switching between different magnetospheric ``modes'', with different
radio, X-ray, and $\gamma$-ray luminosities and different spin-down
rates \citep{Lyne71,Herm13,Alla13}. It is therefore possible that PSR
J1023+0038 is still active as a radio pulsar but may have entered a
different magnetospheric configuration. Since optical observations
make it clear that PSR J1023+0038 acquired a disk at about the same
time the radio, X-ray, and $\gamma$-ray properties changed, Occam's
razor suggests that if PSR J1023+0038 has switched modes it must be
linked to the accretion, presumably triggered by the injection of
small amounts of hadronic material into the light cylinder, although,
to our knowledge, no such mechanism has been proposed.

Energy considerations do not rule out such a scenario: assuming that
the spin-down rate has not changed substantially, that would imply
that PSR J1023+0038's X-ray efficiency was 7\% (in the high mode, and
neglecting any X-ray emission from the disk) and its $\gamma$-ray
efficiency was 15\%; neither number exceeds 100\%. The flares can
exceed the pulsar's spin-down luminosity, so presumably these would
have to originate in the disk, although this appears to conflict with
the fact that the pulsations are suppressed during flares.
Accretion-induced mode switching would explain the rapid switching
between relatively stable X-ray modes, though as the low mode is still
substantially brighter than the radio pulsar's X-ray emission, we
would have to postulate at least two distinct accretion-induced modes.

Accretion-induced mode switching also provides an explanation for the
pulsar's disappearance in radio, although no radio MSP has been
observed to null in radio and their broad beaming (in particular PSR
J1023+0038's broad radio profile) makes it unlikely a reconfigured
beam could miss the Earth.  If the radio pulsar mechanism is active,
it should be producing a pulsar wind, and it becomes difficult to see
how a disk can remain in the system: the pulsar wind pressure falls
off like radiation pressure, but in all disk models, the ram pressure
falls off more rapidly so no stable balance can exist outside the
light cylinder \citep[][]{Shva70}, while substantial amounts of
material inside the light cylinder would be expected to short out the
pulsar mechanism completely.

Overall, we deem accretion-induced mode switching highly unlikely but
cannot definitively rule it out. A detection of $\gamma$-ray
pulsations in the current state would strongly support such an
explanation.  We are carrying out monitoring campaigns with Arecibo,
the Lovell, and the WSRT. These will time the pulsar as soon as it
re-activates in radio, which will measure a mean spin-down during the
accretion-disk state, possibly ruling out accretion-induced mode
switching.

\subsection{Comparison with Other Systems}

\subsubsection{XSS J12270--4859}
In 2012 November or December, the peculiar nearby LMXB and bright
\textit{Fermi} LAT source XSS J12270--4859 (1RXS J122758.8--485343)
underwent a substantial decline in optical/X-ray brightness. Follow-up
optical, X-ray, and radio observations revealed that the accretion
disk had disappeared and the 1.69 ms radio pulsar has switched on
\citep{Bassa14,Bog14b,Roy14}. These findings have established that XSS
J12270--4859 is a close analog to PSR J1023+0038 and only the third
system seen to undergo a MSP to LMXB (or vice-versa) state
transformation.

In its LMXB state, XSS J12270--4859 exhibited a rapid variablity
pattern similar to what we observe in PSR J1023+0038
\citep{Saitou09,deM10,deM13}. Specifically, the same rapid switches
between two flux modes and occasional intense flares are seen.  This
implies that in the LMXB state the same processes operate in both
systems, which was recently confirmed with the detection of X-ray
pulsations in the high mode of this object \citep{Pap14b}.

The drops to the low mode had ingress and egress timescales of
$\sim$10 s and low mode durations between 200 and 800 s. These
transitions were observed in the X-ray and near-UV bands but were
absent in the ground-based optical observations, suggesting an origin
close to the neutron star.  \citet{deM10} attributed the dips
occurring immediately after flares to a rapid episode of accretion
onto the neutron star (corresponding to the flare) and the
corresponding emptying of a reservoir of accreting material (the dip)
and subsequent filling up of the inner regions of the accretion disk.
In contrast, in the long X-ray exposures of PSR J1023+0038 presented
here, we do not find preceding flaring episodes for all the dips but
there is evidence for dips occuring before the majority of intense
flares.

\subsubsection{PSR J1824--2452I}
The X-ray transient IGR J18245--2452 in M28 contains the first neutron
star observed to switch between rotation-powered and accretion-powered
pulsations \citep{Pap13}. The source exhibited a luminous X-ray
outburst in March 2013 reaching a peak 0.5--10 keV luminosity
of $\sim$$5\times10^{36}$ erg s$^{-1}$ \citep[based on \textit{Swift}
  XRT spectra;][]{Lin14a}.  \citet{Pap13} discovered 254 Hz X-ray
pulsations during two \textit{XMM-Newton} observations taken on 2013
April~3 and 13. Remarkably, the spin frequency was identical to that
of a previously known radio MSP, PSR J1824--2452I. The system was
detected again as a radio MSP after the outburst.

In an archival 200-ks \textit{Chandra} ACIS-S observation of PSR
J1824--2452I in 2008 August \citep{Pap13,Lin14a} the
system was at a luminosity level of $\sim$$10^{33}$ erg s$^{-1}$ and
also exhibited large-amplitude variability. While no strong flares
were observed, one low flux mode flux interval lasted for nearly 10
hours instead of several minutes like in PSR J1023+0038.  The
transitions between the two flux states in PSR J1824--2452I occurred
on time-scales of $\sim$500 s instead of $\sim$10 s. The reason for
these substantial differences is not known but probably depends on the
particular combination of pulsar and binary parameters and the
accretion rate.

\subsubsection{SAX J1808.4--3658}
The transient X-ray source SAX J1808.4--3658 is the first accreting
neutron star binary from which coherent millisecond X-ray pulsations
were detected \citep{Wij98}. In its quiescent state, \citet{Camp02}
and \citet{Heinke09} have found 0.5–-10 keV luminosities of
$(8-9)\times10^{31}$\,erg s$^{-1}$ assuming a distance of 3.5 kpc.  At
this level SAX J1808.4--3658 exhibits an X-ray spectrum that is
well-fitted by an absorbed power-law ($\Gamma\approx 1.7-1.8$) with no
requirement for a thermal component.

During the return from the 2005 outburst state to quiescence,
long-term \textit{Swift} monitoring revealed that SAX J1808.4--3658
alternated between two different luminosity levels,
$\approx$$1\times10^{33}$ and $\approx$$3\times10^{32}$ erg s$^{-1}$
(for $D=3.5$ kpc), in observations separated by $\sim$days
\citep{Camp08}. These luminosity levels are comparable to the high and
low modes seen in PSR J1023+0038, XSS J12270--4859, and J1824--2452I,
suggesting that SAX J1808.4--3658 possibly experienced the same moding
behavior during this period (although due to the limited photon
statistics this cannot be firmly verified).  If we further extend the
analogy with these systems, the true quiescent level of SAX
J1808.4--3658 at $\lesssim$$10^{32}$\,erg s$^{-1}$ possibly
corresponds to the disk-free radio pulsar state \citep{Homer01} in
which an intra-binary shock dominates the X-ray emission, as found PSR
J1023+0038 and analogous systems
\citep[see][]{Arch10,Bog05,Bog11,Bog14a}.

\subsubsection{Centaurus X-4}
\citet{Cha14} conducted joint \textit{NuSTAR} and
\textit{XMM-Newton} observations of the long-known, nearby LMXB Cen
X-4.  Although at a comparable luminosity level, unlike PSR
J1023+0038, the non-thermal component of Cen X-4 in quiescence shows a
spectral cutoff at $\sim$10 keV. As shown in \citet{Ten14}, for PSR
J1023+0038, the non-thermal emission is present at least up to
$\sim$30 keV. This implies that the non-thermal component in the
spectrum of Cen X-4 may be of an entirely different origin than what
is seen in PSR J1023+0038. In addition, Cen X-4 exhibits a dominant
thermal emission component, presumably due to radiation from most of
the neutron star sufrace.  In contrast, PSR J1023+0038 does not show a
prominent thermal component, an indication that the accretion flow
does not induce substantial heating of the neutron star. The possible
presence of a faint thermal component in the high flux mode of PSR
J1023+0038 indicates that the polar caps are heated only superficially
and the deposited heat is quickly reradiated.

The marked differences in observed properties between the two systems
might be ascribed to a very weakly magnetized compact object in Cen
X-4, such that the accretion flow can proceed unimpeded down to the NS
surface \citep[see, e.g.,][for further details]{DAng14}.

\subsection{Previous Interpretations}
Prior to results from \citet{Arch14} and from the analysis presented
here, although the presence of an accretion disk was undisputed, it
was unclear whether the pulsar wind in the PSR J1023+0038 binary was
still active or if material was actually flowing down to the neutron
star surface, even intermittently.  Several interpretations have been
offered to account for the observed behavior.

\citet{Lin14a} proposed that this bi-stable X-ray flux switching in
PSR J1824--2452I is due to rapid transitions between magnetospheric
accretion (high mode) and pulsar wind shock emission (low mode)
regimes, with the two different non-thermal emission mechanisms
coincidentally producing the same power-law spectrum with
$\Gamma\approx1.7$.

An alternative interpretation was offered by \citet{Pap14} to account
for the properties of XSS J12270--4859 in its LMXB state, which is
fully applicable to PSR J1023+0038 as well. The mode switches were
attributed to rapid emptying and refilling of the inner accretion disk
due to the action of the pulsar wind or the propeller mechanism, with
little to no accretion taking place.

\citet{Stap14}, \citet{Tak14}, and \citet{Cot14} have provided yet
another explanation for the multi-wavelength phenomenology of PSR
J1023+0038. In their model, the pulsar wind is still active but the
radio pulsations are dispersed by evaporating material from the
accretion disk or material that is engulfing the pulsar. The great
increase in X-ray luminosity during the accreting state compared to
the radio pulsar state is then the result of a stronger intra-binary
shock since the pulsar wind encounters a stronger outflow from the
companion that is much closer to the pulsar.  The enhancement in
$\gamma$-ray emission can be accounted for by inverse Compton
scattering of UV emission by particles from the pulsar wind.

Based on the wealth of observational information we have presented
here, we can evaluate the viability of the interpretations described
above.  Any self-consistent explanation for the behavior of PSR
J1023+0038, and by extension quite likely XSS J12270--4859 and PSR
J1824--2452I as well, needs to account for the large-amplitude X-ray
variability, especially the intermittent, rapid ($\sim$10 sec)
switching between two discrete flux levels, the coherent X-ray
pulsations during the high flux mode, and the flat-spectrum radio
emission.  The interpretations that invoke an active pulsar wind
during the high flux mode in particular are difficult to reconcile
with the observed X-ray pulsations and flat radio spectrum, which
leaves accretion processes as the more probable explanation for this
emission.

\subsection{The Accretion Physics of PSR J1023+0038}
X-ray binaries containing neutron stars are routinely studied in great
detail when actively accreting ($\gtrsim$$10^{35}$ erg s$^{-1}$) and
most of our understanding of the physics of accretion onto magnetized
neutron stars comes from such instances.  Specifically, it is
well-established that at high luminosities matter flows down to the
neutron star surface, as evidenced by the detection of thermonuclear
bursts in LMXBs and X-ray pulsations in accreting millisecond X-ray
pulsars \citep[AMXPs; see, e.g.,][and references therein]{Pat12}.
However, none of the known AMXPs are close enough to Earth to easily
establish whether channeled accretion also reaches the neutron star
during quiescence.  The discovery of coherent X-ray pulsations from
PSR J1023+0038 at a luminosity level of $\sim$$10^{33}$ erg s$^{-1}$
provides crucial information regarding the physical processes that
operate in quiescence.

In basic accretion models onto magnetized stars, the accretion disk
has a truncation at the magnetospheric radius, where the ram pressure
of the infalling material balances the magnetic pressure of the field
\citep{Pringle72}. This occurs at a distance $r_m\simeq
(BR^3)^{4/7}(GM/2)^{1/7}\dot{M}^{-2/7}$. If this magnetospheric radius
exceeds the co-rotation radius, $r_c=GM/\Omega^2$, where the Keplerian
orbital velocity of the accretion flow equals the rotation rate of the
neutron star, the system enters the ``propeller'' regime
\citep{Ill75}. The resulting centrifugal barrier created by the
rapidly-spinning neutron star does not permit accretion and expels the
infalling material from the system.  For low enough mass inflow rates,
the radio pulsar mechanism might also activate and sweep the inflowing
matter away. Based on this, such models predict that accretion onto
the neutron star surface is an unlikely mechanism for producing the
quiescent emission of AMXPs, as it requires very low magnetic fields
and/or long spin periods. However, our findings for PSR J1023+0038
suggest that accretion onto the neutron star surface does occur
despite the remarkably low implied accretion rate and rapid spin of
the star, which is also sufficiently magnetized to force channeled
accretion.

If the X-ray luminosity observed in the high flux mode is generated
entirely due to the liberation of gravitational potential energy of
the material landing on the stellar sufrace, the implied rate is
$\sim$$9\times10^{-13}$ M$_{\odot}$ yr$^{-1}$, corresponding to just
$\sim$$10^{-5}-10^{-4}L_{\rm Edd}$.  It remains unclear, however, what
portion of the gas flowing through the disk actually reaches the star
so this value only provides a lower limit on the accretion rate at
$\sim$$r_c$.

The observed X-ray luminosity of the accretion disk depends on
  three factors: (i) the mass accretion rate through the inner edge of
  the disk; (ii) the location of disk inner edge; (iii) the radiative
  efficiency of the accretion flow. For the case of a strong
  propeller, only a small portion of matter falls onto the star. The
  inner disk will thus have $\dot{M}_{in}$ $\sim$100 times larger than
  what is accreted onto the star ($\dot{M}_{NS}$). As a result, even
  though the disk is truncated it should still generate a fairly high
  X-ray luminosity, which is not observed in PSR
  J1023+0038. Therefore, for the strong propeller interpretation to be
  viable it is necessary to invoke a radiatively ineficient accretion
  flow \citep{Rees82}.  In that case, the disk changes from being
optically thick to being optically thin but geometrically thick, and
the gravitational potential energy of the material is not emitted as
radiation. A sufficiently high accretion rate can result in
$r_m\approx r_c$ such that a small fraction of the material can reach
the stellar surface.

For a ``weak'' propeller, $\dot{M}_{in} \approx \dot{M}_{NS}$,
  resulting in comparable X-ray luminosities.  If we assume that in
  the low mode no accretion takes place onto the neutron star surface,
  the observed X-ray luminosity ($5.1\times10^{32}$ erg s$^{-1}$) would then
  corresponds to $\dot{M}_{in}$. For $\dot{M}_{in} =
  \dot{M}_{NS}$, the high/low mode luminosity ratio, corrected for
  gravitational redhsift, implies a truncation radius in the
  range $\sim$$80-150$ km, depending on the choice of stellar mass and
  radius. For reference, the light cylinder radius of PSR J1023+0038
  is $cP/2\pi=81$ km, suggesting that the accretion flow clears out of the
  pulsar magnetosphere during the low mode.  In this weak propeller
scenario, some mechanism still needs to account for the intermittent
but very steady episodes of accretion (corresponding to the high flux
mode).

A possible interpretation for the X-ray mode switching
involves a variation of the so-called ``dead'' disk
\citep{Sun77,Spru93}, where gas is accumulated near co-rotation, which
exerts inward pressure to balance the magnetic stress. The transitions
between the high and low modes can be interpreted as being due to the
rapid emptying and refilling of this reservoir of trapped material
\citep[see, e.g.,][]{DAng10,DAng12}.  Although the cyclic variation
between high and low accretion rate is a feature of this ``trapped''
disk model, the manner in which this occurs is not consistent with the
``square-wave'' X-ray mode switching observed in PSR J1023+0038,
especially the remarkably steady luminosity levels in both the high
and low modes.  The oscillation timescales in the trapped disk model
are much shorter (up to $\sim$100--1000 s) than the variability
observed in PSR J1023+0038. In addition, flare-like spikes in
accretion rate are seen to occur in simulations \citep{Lii14}
-- the disk goes through cycles of building up gas and then dumping it
onto the star, with the build-up timescales generally being longer
than the dumping timescales.

Throughout the discussion above, it is assumed that the gas can
effectively couple to the magnetic field. However, the magnetic
field can become disconnected from the disc from differential rotation
between the disc and the star \citep{Love95}, at which point it is not
completely clear how efficiently the gas reconnects the field lines to
accrete onto the magnetic pole(s) of the star.  If there is perfect
coupling, in the propeller regime no matter is is able to accrete at
low $\dot{M}$.  However, simulations (which are ideal MHD with
numerical diffusion) show that some accretion through the centrifugal
barrier can occur because the gas is disconnected from the
fast-spinning magnetosphere \citep[see, e.g.,][]{Rom04,Lii14}. In
\citet{DAng14}, the results of these simulations were applied to this
modified propeller -- most of the gas is expelled but a small fraction
is accreted, and this fraction could, in principle, produce coherent
X-ray pulsations.

A qualitative description for the moding behavior might involve a
transition between the trapped disk and propeller mode. A disk
may remain trapped because of extra gas in the inner region of the
disk (compared with a normal accretion disk), which absorbs the
angular momentum from the disk-field interaction. This state could
co-exist with an outflow of gas, which could then erode enough gas in
the disk so that the system enters the propeller regime, and no
gas accretes onto the star, hence it becomes much dimmer. The reverse
could occur as well. If not all the material around $r_{\rm m}$ is
expelled in an outflow, gas begins to build up into a trapped disk
again. Given sufficient build up of matter, accretion on to the star
can commence again, which will result in a rapid jump in
luminosity. However, detailed modeling is necessary to assess the
viability of this hybrid trapped disk -- propeller mode scenario.

One of the most notable features of the high X-ray flux mode in PSR
J1023+0038 is that, despite its intermittent nature, the luminosity
and pulse shape are highly reproducible and stable on time scales up
to at least seven months (corresponding to the separation between the
two \textit{XMM-Newton} observations). In addition, the duty cycle of
the high mode is $\sim$70\%, meaning that this steady accretion occurs
most of the time.  Any accretion model that requires an instability to
instigate accretion onto the neutron star would thus have difficulties
explaining this steady behavior because the instability would result
in brief episodes of accretion and much more erratic burst-like flux
variability.

In this sense, the occasional flares observed in our X-ray data are
more consistent with being due to accretion flow instabilities. For
instance, \citet{Kul08} have found that Rayleigh-Taylor instabilities
may produce ``tongues'' of plasma that break through the magnetosphere
and reach the stellar surface close to the equator.  The shape and
number of the tongues changes with time on the inner disc dynamical
time-scale.  The shape, number and location of the resulting hot spots
varies on timescales comparable to the dynamical time scale of the
inner disk.  This erratic behavior can, in principle, account for the
absence of coherent X-ray pulsations during the flare mode of PSR
J1023+0038.  However, the instability appears for cases when the
angles between the spin and magnetic axes are
$\theta\lesssim$$30^{\circ}$. The double-peaked X-ray pulse profiles
of PSR J1023+0038 suggests a greater misalignment of the two axes for
this neutron star.  In addition, as per \citet{Kul08}, for neutron
stars such instabilities occur for accretion rates above
$\dot{M}_{*,\rm crit} \approx 2.2\times10^{-9} {\rm M}_{\odot} {\rm
  yr}^{-1} (B/10^9\,{\rm G})^2 (R_{NS}/10\,{\rm km})^{5/2}
(M_{NS}/1.4\,{\rm M}_{\odot})^{-1/2}$. For the magnetic field strength
of $9.7\times10^7$ G deduced from radio timing \citep{Arch13},
$M_{NS}=1.4$ M$_{\odot}$, and $R_{NS}=10$ km, for PSR J1023+0038 we
obtain $\dot{M}_{*,\rm crit}\approx 2\times 10^{-11}$ M$_{\odot}$
yr$^{-1}$, an order of magnitude greater than implied by the flare
luminosity.  Finally, the timescales for the instability are much
shorter than the variability we observe in PSR J1023+0038; they are
typically of order a few times the spin period instead of tens of
seconds.  In light of these differences, although the basic
phenomenology seems to fit fairly well, especially the absence of
pulsations in the flare mode, it is not clear whether this instability
model is applicable to the flare mode of PSR J1023+0038.

It is apparent that further theoretical efforts are required to
establish which, if any, existing models of accretion can adequately
explain the observed behavior of PSR J1023+0038. Specifically, it is
necessary to consider a process that ejects a large portion of the accretion
flow material while maintaining ``quiescent'' level X-ray
luminosities. The remaining material needs to be able to reach the
neutron star polar caps via a steady channeled flow that is restricted
to a very narrow range of accretion rates. This flow undergoes
occasional interruptions, manifested by abrupt changes to a similarly
stable but lower luminosity mode. This pattern is interspersed with
short-lived episodes of higher X-ray luminosity, presumably due to
unstable accretion.

\section{CONCLUSIONS} 
We have presented extensive multi-wavelength observations of PSR
J1023+0038 in its LMXB state. The wealth of data offer unique insight
into the behavior of accreting MSPs in the quiescent regime.  The
system appears to exhibit short-lived but very frequent and
exceptionally stable episodes of chanelled accretion onto the stellar
magnetic poles. This activity appears to be limited to a remarkably
narrow luminosity range around $\sim$$3\times10^{33}$ erg
s$^{-1}$. The frequent and rapid switches to a second, lower
luminosity mode with $\sim$$5\times10^{32}$ erg s$^{-1}$ during which
no pulsations are observed, can be interpreted as being non-accreting
intervals.  At least a portion of the occasional X-ray/optical flares
may be produced by enhanced, spasmodic accretion due to some form of
disk instability or deformation of the magnetosphere.  Although
certain propeller and trapped disk models predict qualitatively similar
behavior, none are fully consistent with the observed properties of
PSR J1023+0038, especially the accretion-induced pulsations produced
at a luminosity $\sim$100 times lower than previosuly observed in an
AMXP.

The episodic reappearance of an accretion disk in PSR J1023+0038
implies that the current disk will eventually recede and the binary
will revert to its radio pulsar state yet again. In such an event, it
is important to identify any differences in the system between the pre
and post LMXB state periods.

Further insight into this peculiar system can be gained with
additional observations, especially contemporaneous X-ray and
continuum radio observations over a long time span, to conclusively
establish any correlated variability between the two bands.  Perhaps
most importantly, it is crucial to monitor the long-term spin behavior
of PSR J1023+0038 in order to establish what kinds of torques are
being imparted onto the neutron star, which may have profound
implications for understanding of accretion onto highly magnetized
objects, in general.

\acknowledgements 
A.M.A.~and J.W.T.H.~acknowledge support from a
Vrije Competitie grant from NWO.  A.T.D. acknowledges support from an
NWO Veni Fellowship. J.W.T.H. and A.P. acknowledge support from NWO
Vidi grants. J.W.T.H.~also acknowledges funding from an ERC Starting
Grant ``DRAGNET'' (337062).  A portion of the results presented was
based on observations obtained with \textit{XMM-Newton}, an ESA
science mission with instruments and contributions directly funded by
ESA Member States and NASA.  This work was based in part on
observations obtained at the MDM Observatory, operated by Dartmouth
College, Columbia University, Ohio State University, Ohio University,
and the University of Michigan. This research is based in part on
observations made with ESO Telescopes at the Paranal Observatory under
programme ID 292-5011.  The WSRT is operated by ASTRON (Netherlands
Institute for Radio Astronomy) with support from the Netherlands
Foundation for Scientific Research. Access to the Lovell Telescope is
supported through an STFC consolidated grant. The National Radio
Astronomy Observatory is a facility of the National Science Foundation
operated under cooperative agreement by Associated Universities, Inc.
The EVN (\url{http://www.evlbi.org}) is a joint facility of European,
Chinese, South African, and other radio astronomy institutes funded by
their national research councils.  LOFAR, the Low Frequency Array
designed and constructed by ASTRON, has facilities in several
countries, that are owned by various parties (each with their own
funding sources), and that are collectively operated by the
International LOFAR Telescope (ILT) foundation under a joint
scientific policy.  This research has made use of the NASA
Astrophysics Data System (ADS).

Facilities: \textit{XMM, Swift, ESO/VLT}

\end{document}